\documentclass[reprint, amsmath,amssymb, aps, prd]{revtex4-2}

\usepackage{graphicx}% Include figure files
\usepackage{dcolumn}% Align table columns on decimal point
\usepackage{bm}% bold math
\usepackage{upgreek}% bold math
\usepackage{hyperref}
\usepackage{xspace} 
\usepackage{dsfont}

\usepackage{pennames}

\usepackage{xstring}
\makeatletter
\AtBeginDocument{%
  \def\doi#1{\href{https://doi.org/#1}{doi:\StrSubstitute{#1}{_}{\_}}}}
\makeatother
\newcommand{\WZ}{\ensuremath{\mathrm{W}\mathrm{Z}}\xspace} 
\newcommand{\ttbar}{\ensuremath{\mathrm{t}\bar{\mathrm{t}}}\xspace}
\newcommand{\ttA}{\ensuremath{\mathrm{t}\bar{\mathrm{t}}\gamma}\xspace}
\newcommand{\CP}{\ensuremath{\textit{CP}}\xspace}
\newcommand{\pt}{\ensuremath{p_{\mathrm{T}}}\xspace}

\begin{document}

%\preprint{APS/123-QED}

\title{Equivariant neural networks for robust \CP observables}% Force line breaks with 

\author{Sergio S\'anchez Cruz}
\email{sergio.sanchez.cruz@cern.ch}
\affiliation{European Organization for Nuclear Research (CERN), 1211 Geneva 23, Switzerland}

\author{Marina Kolosova}
\affiliation{University of Florida, Gainesville, Florida, USA}

\author{Clara Ram\'on \'Alvarez}
\affiliation{Universidad de Oviedo and ICTEA, Oviedo, Asturias, Spain}

\author{Giovanni Petrucciani}
\affiliation{European Organization for Nuclear Research (CERN), 1211 Geneva 23, Switzerland}

\author{Pietro Vischia}
\affiliation{Universidad de Oviedo and ICTEA, Oviedo, Asturias, Spain}

\date{\today}% It is always \today, today,
             %  but any date may be explicitly specified

\begin{abstract}

We introduce the usage of equivariant neural networks in the search for violations of the charge-parity (\CP) symmetry in particle interactions at the CERN Large Hadron Collider. We design neural networks that take as inputs kinematic information of recorded events and that transform equivariantly under the a symmetry group related to the \CP transformation. We show that this algorithm allows one to define observables reflecting the properties of the \CP symmetry,  showcasing its performance in several reference processes in top quark and electroweak physics. Imposing equivariance as an inductive bias in the algorithm improves the numerical convergence properties with respect to other methods that do not rely on equivariance and allows one to construct optimal observables that significantly improve the state-of-the-art methodology in the searches considered.

\end{abstract}
\maketitle

\section{Introduction}
\label{sec:intro}

The violation of the charge-parity (\CP) symmetry is one of the necessary conditions to allow baryogenesis~\cite{Sakharov:1967dj} in the early Universe and, therefore, one of the key ingredients needed to explain the observable Universe. The only source of \CP violation in the standard model (SM) of particle physics is introduced as an extension of the Cabibbo mixing mechanism~\cite{PhysRevLett.10.531} by the Kobayashi-Maskawa (KM) mechanism in the electroweak sector. Several observations of \CP violation stemming from the KM mechanism have been performed in the past years~\cite{ParticleDataGroup:2022pth}, but this mechanism alone cannot explain the magnitude of the matter-antimatter asymmetry present in the universe. In contrast, numerous extensions of the SM incorporate additional sources of \CP violation. This, together with the fact that \CP violation manifests itself as a striking experimental signature, makes searches for \CP violation one of the most interesting probes for physics beyond the SM (BSM).

In this manuscript, we introduce a novel technique to enhance searches for \CP-violating phenomena. We will focus in collider experiment searches in the SM effective field theory (SMEFT) framework: we, however, highlight that the techniques described can be applied to other contexts in the search for \CP violation. 

In the SMEFT, new physics contributions are introduced as a set of operators $\mathcal{O}_{i,d}$ that are added to the SM Lagrangian density weighted by coefficients, denoted Wilson coefficients (WCs) $c_{i,d}$, that regulate the size of the contribution of each operator:

\begin{equation}
 \mathcal{L}_{\mathrm{EFT}} = \mathcal{L}_{\mathrm{SM}} + \sum_{i,d}  \frac{c_{i,d}}{\Lambda^{d-4} } \mathcal{O}_{i,d}\,.
 \end{equation}

Operators in the SMEFT are hierarchically sorted by their natural dimension $d$, starting at dimension six for scenarios relevant to our studies~\cite{Grzadkowski:2010es}. At dimension six, there are 1350 \CP-invariant operators and 1149 \CP-odd operators~\cite{Alonso:2013hga}, assuming three generations. These operators introduce additional interactions to those present in the SM and, from the phenomenological standpoint, we distinguish between two BSM contributions to any experimental observable e.g. a given process' cross section: we denote ``linear'' contributions those that arise from the interference between the SM and BSM Feynman diagrams, which contribute linearly as a function of the WCs, and ``quadratic'' as those that stem exclusively from BSM diagrams, which contribute quadratically as a function of the WCs. In the SMEFT, the linear contribution is the only one that does not receive contributions from operators with $d>6$; hence, these contributions would be an unambiguous indication of dimension-six effects. 

In addition, when considering \CP-odd operators, linear terms are particularly interesting from the experimentalist's point of view. While the quadratic contributions induced by these operators are \CP-invariant, the linear contributions are odd under \CP transformations and, therefore, manifest themselves as asymmetries in \CP-odd observables. Searches based on this feature are particularly robust against systematic uncertainties such as those related to the modeling of the SM prediction, as those are predominantly \CP-invariant, and, therefore, cannot induce an asymmetry in a \CP-odd observable. 

In this paper, we exploit machine learning techniques to build optimal \CP-odd observables to search for new physics, using neural networks that transform equivariantly with a symmetry group associated with the \CP transformation. Machine learning is widespread in the context of high-energy physics; in particular, in the context of EFT searches, numerous techniques have been developed to maximally extract information from the Large Hadron Collider (LHC) data, using dense neural networks, graph neural networks and others (see Refs.~\cite{hepmllivingreview,Calafiura:2022ges} for reviews). Several studies~\cite{Satorras2021EnEG,Villar:2021wnx,Dolan:2020qkr,Favoni:2020reg,Bulusu:2021njs,Gong:2022lye,Bogatskiy:2022hub,Favoni:2022mcg,Bogatskiy:2022czk,Buhmann:2023pmh,Murnane:2023kfm,Bogatskiy:2023nnw} have also shown the power of equivariant neural networks in exploiting physical symmetries present in a number of problems. 

This paper is organized as follows. Section~\ref{sec:algo} describes the algorithm we have developed. Section~\ref{sec:application} describes the application of the algorithm to searches for \CP violation in \ttbar, \WZ and \ttA production, showcasing its most interesting properties. We close the document with some conclusions in section~\ref{sec:conclusions}.

\section{Algorithm and properties} 
\label{sec:algo} 

Our algorithm constructs a function $f: \mathcal{D}\to\mathbb{R}^{n_1+n_2}$, where the domain $\mathcal{D}$ is the space of per-event input features. We note that this space can be a subset of $\mathbb{R}^m$ for a fixed dimension $m$, representing e.g. the four-momenta of a fixed set of objects in the event, but can also be as general as a point cloud including the variable-size list of objects in the event. We  will design $f$ in a way that the first $n_1$ dimensions of its score are odd under \CP transformation of its input and the latter $n_2$ dimensions are invariant under such transformations. For the purpose of constructing  a single \CP-odd observable, one chooses $n_1=1$ and $n_2=0$; however, the algorithm we present is more general than that. Building an algorithm with $n_1 > 1$  can be used to design a set of observables that are sensitive to a given set of $n_1 > 0$ \CP-odd operators.  In addition, algorithms with $n_2 > 0$ are appropriate when building discriminators aiming to discriminate among different SM backgrounds or when attempting to obtain variables sensitive to \CP-even operators or the quadratic term induced by a \CP-odd operator. 

We impose that $f$ transforms equivariantly with respect to the $\mathcal{Z}_2$ symmetry group, under the following representations in the domain and target spaces. For the domain, we choose the representation given by $\lbrace \mathds{1}, h_{\CP} \rbrace$, where $\mathds{1}$ denotes the neutral element ($\mathds{1}(d)=d$ for all $d$ in $\mathcal{D}$) and $h_{\CP}$ the \CP transformation applied to the input variables, which needs to be specified for each specific problem to be solved. For the target space, we choose $\lbrace \mathds{1}_{n_1+n_2}, \tilde{h}_{\CP}\rbrace$, where $\mathds{1}_{n_1+n_2}$ is the identity matrix in $\mathbb{R}^{n_1+n_2}$ and $\tilde{h}_{\CP}$ is a transformation that acts on a given element of $\mathbb{R}^{n_1+n_2}$, $(x_1,...,x_{n_1+n_2})$, as $\tilde{h}_{\CP}(x_1,...,x_{n_1+n_2})=(-x_1,..., -x_{n_1}, x_{n_1+1},...,x_{n_1+n_2})$. The most general function satisfying this equivariant property is defined by its action on any element $d$ in $\mathcal{D}$:

\begin{align}
    \label{eq:masterformula}
    f( d )  =   & \left( g_1( d ) - g_1( h_{\CP}(d)), \right. \nonumber \\ 
    & \ldots,  \nonumber \\
     & g_{n_1}( d ) - g_{n_1}( h_{\CP}(d)), \nonumber \\ 
     & g_{n_1+1}( d ) + g_{n_1+1}( h_{\CP}(d)),   \nonumber \\ 
     & \ldots, \nonumber \\ 
     & \left.  g_{n_1+n_2}( d ) + g_{n_1+n_2}( h_{\CP}(d)) \right)\,,
\end{align}

where $g_i$ are arbitrary functions $g_i: \mathcal{D} \to \mathbb{R}$. It is straightforward to prove that the first $n_1$ components of $f$  in equation~\eqref{eq:masterformula} are \CP-odd, and the last $n_2$ are \CP-even. In addition, any function fulfilling this equivariant condition can be written under the functional form of Eq.~\ref{eq:masterformula}. In practice, $g_i$ can be parametrized by an appropriately designed neural network and, when doing so, such a function $f$ can be used to approximate any function satisfying the equivariant property described above. In this paper, we parametrize $g_i$ using multilayer perceptrons (MLPs) with four hidden layers and between 20 and 80 neurons each, using leaky rectified linear units as activation functions. We highlight that more advanced architectures could be used for more complex use cases.

This algorithm allows to build optimal \CP-odd or \CP-invariant observables by training $f$ to minimize the category cross-entropy in a multiclass classification problem~\cite{Bhardwaj:2021ujv} or by building a surrogate model of the likelihood ratio, using any of the loss functions described in~\cite{Cranmer:2015bka,Brehmer:2018kdj,Brehmer:2018eca,Brehmer:2018hga,Brehmer:2019xox,Brehmer:2019gmn,Butter:2021rvz,Chatterjee:2022oco,GomezAmbrosio:2022mpm,Chen:2020mev,Chen:2023ind,Bortolato:2020zcg}, as we do in section~\ref{sec:application}. Our algorithm further extends the method described in Ref.~\cite{Bhardwaj:2021ujv}, which also allows one to build optimal \CP-odd observables, in the following ways. Firstly, our algorithm allows one to naturally introduce an arbitrary number of \CP-odd or \CP-invariant output scores. In addition, our approach imposes the behavior of the model with respect to the \CP asymmetry in the model architecture, rather through the loss function and training strategy, leaving the freedom to choose any cost function suitable for the specific problem. More importantly, the resulting model fulfills the imposed equivariance  properties, regardless of the convergence of the training. This aspect, which is illustrated in Section~\ref{sec:application}, is of utmost importance in searches in the SMEFT framework. Those analyses employ SMEFT predictions by reweighing simulated samples~\cite{Mattelaer:2016gcx} as we describe in Section~\ref{sec:application}. This reweighing procedure reduces the statistical power of the samples and often introduces outliers that may distort the training, spoiling the numerical convergence. In our algorithm, while this effect could induce a nonoptimality of the achieved discriminator, the resulting observables are \CP-odd or invariant by construction, regardless of the convergence of the training. 

\section{Applications of the algorithm} 
\label{sec:application}

We apply our algorithm in the search for \CP violation in the production of a top quark anti-quark pair (\ttbar), the associated production of a $\PW^{\pm}$ and a $\PZ$  boson ($\WZ$) and the associated production of a top quark anti-quark pair with a high-energy photon (\ttA). We employ realistic simulations of these processes at the LHC with a center-of-mass energy of 13 TeV. The hard scattering part of the collision is simulated using the \texttt{MADGRAPH5\_aMC@NLO} generator v2.9.18~\cite{Alwall:2014hca}, including the decays of the top (anti)quarks into $\PW^{\pm}$ bosons and bottom quarks, and the subsequent decays of the $\PW^{\pm}$ and $\PZ$ bosons into leptons~\cite{PhysRevLett.19.1264} or quarks, depending on the final state considered. SMEFT effects are introducing using the \texttt{dim6top}~\cite{Aguilar-Saavedra:2018ksv} model for \ttbar and \ttA production, and the \texttt{SMEFTSim} v3.0~\cite{Brivio:2020onw} model for \WZ production. We use the \texttt{PYTHIA}~v8.3~\cite{Sjostrand:2014zea} package to simulate the parton shower and hadronization, and the \texttt{DELPHES}~v3.5.0~\cite{deFavereau:2013fsa} software with a detector response description based on the ATLAS experiment settings (``ATLAS card''). After the event selection, described in the upcoming sections, we consider samples of 6.6, 5.8, and 2.5 million \ttbar, \WZ and \ttA events, respectively, half of which are used for training and half for testing. In this study, we consider only the effect of these signal processes.

We  obtain a parametric prediction of our detector-level observables as a function of the WCs weighting our simulated events. The weights have a quadratic dependence on the WCs, $w_j (\mathbf{c}) = w_j^{\mathrm{SM}} + \sum_i c_i l_j^i + \sum_{ik} c_i c_k q_{j}^{ik} $, where $j$ labels a given simulated event, $\mathbf{c}$ represents the vector of WCs, and $w_j^{\mathrm{SM}}$, $l_j^i$, and $q_{j}^{ik}$ represent the SM, linear, and quadratic SMEFT contributions. We derive these three quantities for each event using \texttt{MADGRAPH5\_aMC@NLO} with the \texttt{MADWEIGHT}~\cite{Artoisenet:2010cn} module for a sufficiently large of parameter points to derive a quadratic parametrization.

We follow the same procedure used in Ref.~\cite{Chatterjee:2024pbp} to obtain optimal discriminants to capture the linear contribution. This method, which is similar to the SALLY method~\cite{Brehmer:2018kdj}, considers the per-event likelihood ratio between the a given BSM point with WCs $\vec{c}$ and the SM, given by $\frac{p(d|\vec{c})}{p(d|\mathrm{SM})}$, which is a function of the per-event features $d$. Since we are interested in the linear contribution, we consider on the score vector at the SM 

\begin{equation}
    \vec{t}(d) = \nabla_{\vec{c}} \log p(d|\vec{c})|_{\vec{c}=\mathrm{SM}}=\frac{\nabla_{\vec{c}} p(d|\vec{c})|_{\vec{c}=\mathrm{SM}}}{p(d|\mathrm{SM})}\,,
\end{equation} 

which is a sufficient statistic for smaller values of $\vec{c}$, where the linear term dominates. While $p$ is not a tractable quantity as a function of $d$, we can use simulations to learn it. It is indeed shown in Ref~\cite{Chatterjee:2024pbp} that a function $f$ minimizing the loss function:

\begin{equation}
    \label{eq:loss}
    L = \sum_{j\in \mathrm{events}} w_j^{\mathrm{SM}} \left( \frac{l_j^i}{w_j^{\mathrm{SM}}}  -f (d_j) \right)^2\,,
\end{equation}

provides a surrogate of the score vector as a function of the Wilson coefficient $i$ and is, therefore, a sufficient statistic. 

We highlight that this method allows us to naturally account for the contribution from more than one signal processes or background that do not show any EFT dependence. This is achieved by training on a sample of simulated events sampled from all processes of interest. $w_j^{\mathrm{SM}}$ should then incorporate each process' cross section so that the sample contains the relative contribution of each process that is expected in data. 

When considering background events, since they do not show any EFT dependence, $l_j^i$ should be taken to be zero. In this case, the resulting score will play the dual role of, on the one hand, extracting \CP-violating features from the signal and, on the other hand, discriminating between signal events, which will typically take values different from zero, and background events, which will take values closer to zero. 
 
\subsection{CP violation in \texorpdfstring{$\ttbar$}{tt} production} 
\label{sec:ttbar}

We consider \ttbar production in its dileptonic final state. This process could receive contributions from the chromoelectric dipole moment operator $\mathcal{O}_{\mathrm{tG}}^I = \mathrm{Im}\left(  \bar{Q_3} \sigma^{\mu\nu} \mathrm{T}^A u_3 \tilde{\Phi} G_{\mu\nu}^A \right) $, which has been exploited in previous searches~\cite{CMS:2019nrx,CMS:2022voq}. In addition, several observables have been proposed~\cite{Bernreuther:2015yna}, which rely on spin correlations to probe different kinds of \CP-invariant and violating BSM contributions. Therefore, \ttbar production is, together with \WZ production, an excellent test bed for our method and study the interplay with those already-proposed based on analytical studies at parton level.

We select events with two reconstructed leptons (electrons or muons) with transverse momentum $\pt > 15$ GeV and pseudorapidity $|\eta|<2.5$ with the opposite charge. We also require the presence of two jets, clustered with the anti-$k_{\mathrm{T}}$ method~\cite{Cacciari:2008gp} with parameter 0.4, that are separated in $\Delta R = \sqrt{\Delta\phi^2+\Delta\eta^2} < 0.4$ with respect to the selected leptons. We require at least one of the jets to be \Pb-tagged.

\begin{figure*}
    \includegraphics[width=0.24\textwidth]{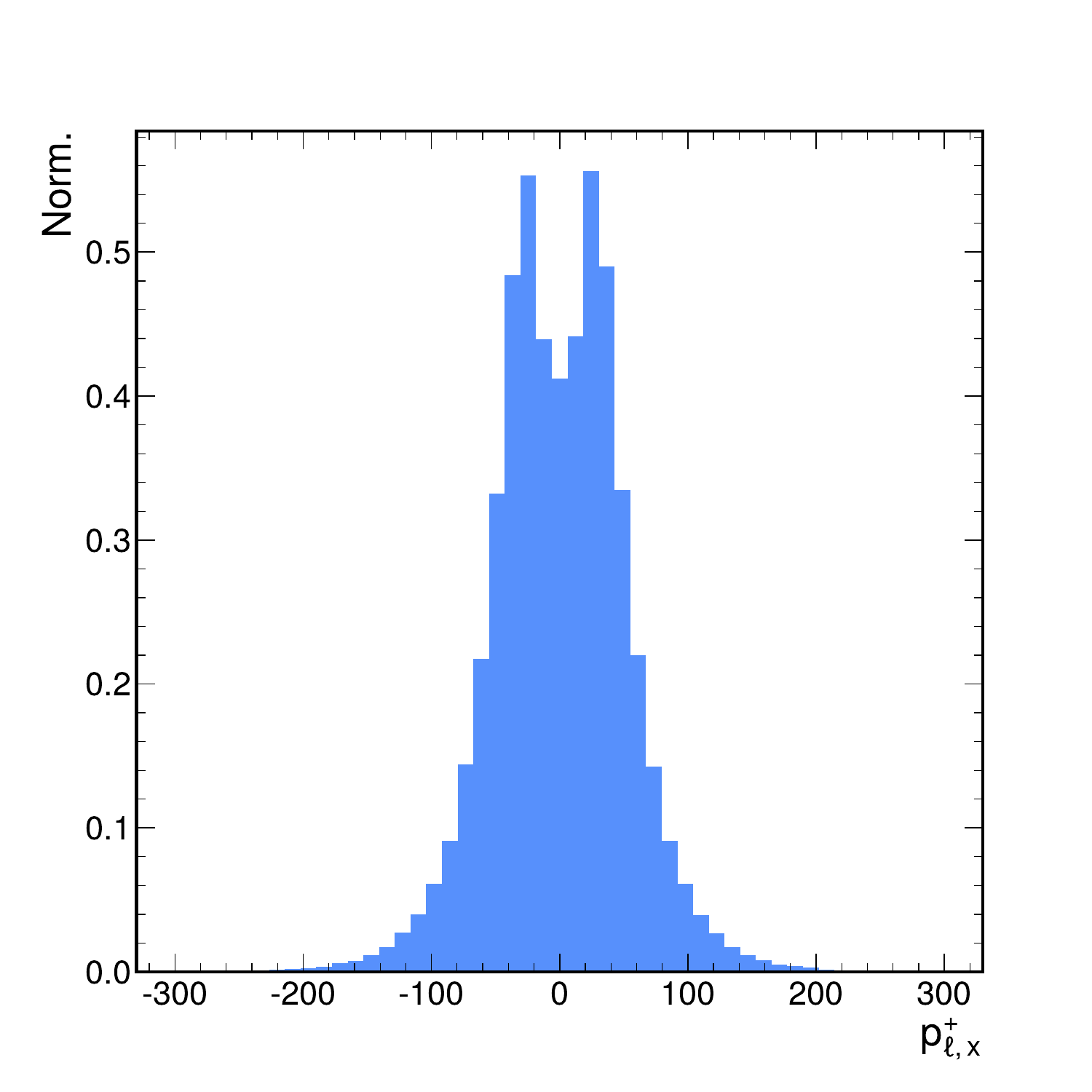}
    \includegraphics[width=0.24\textwidth]{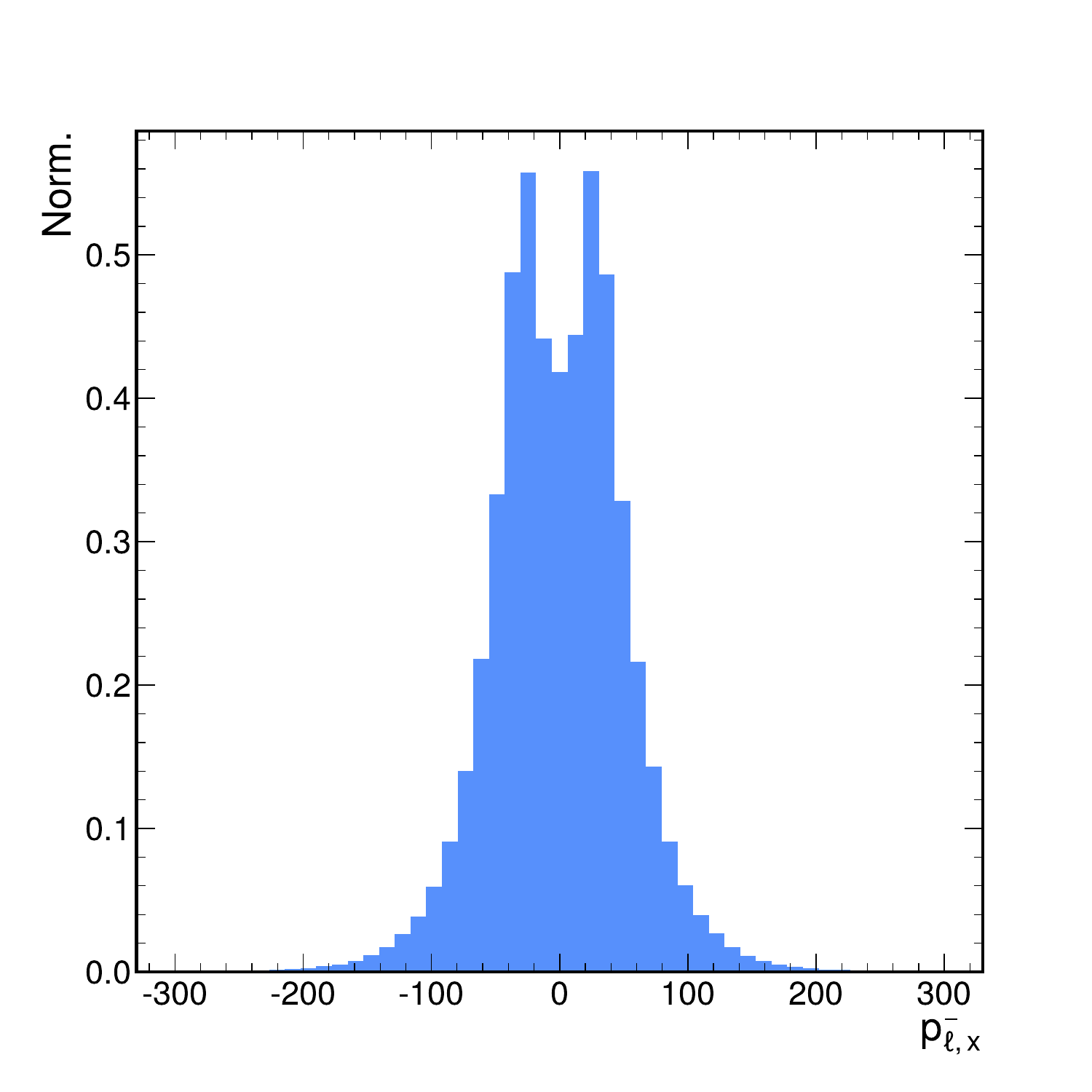}
    \includegraphics[width=0.24\textwidth]{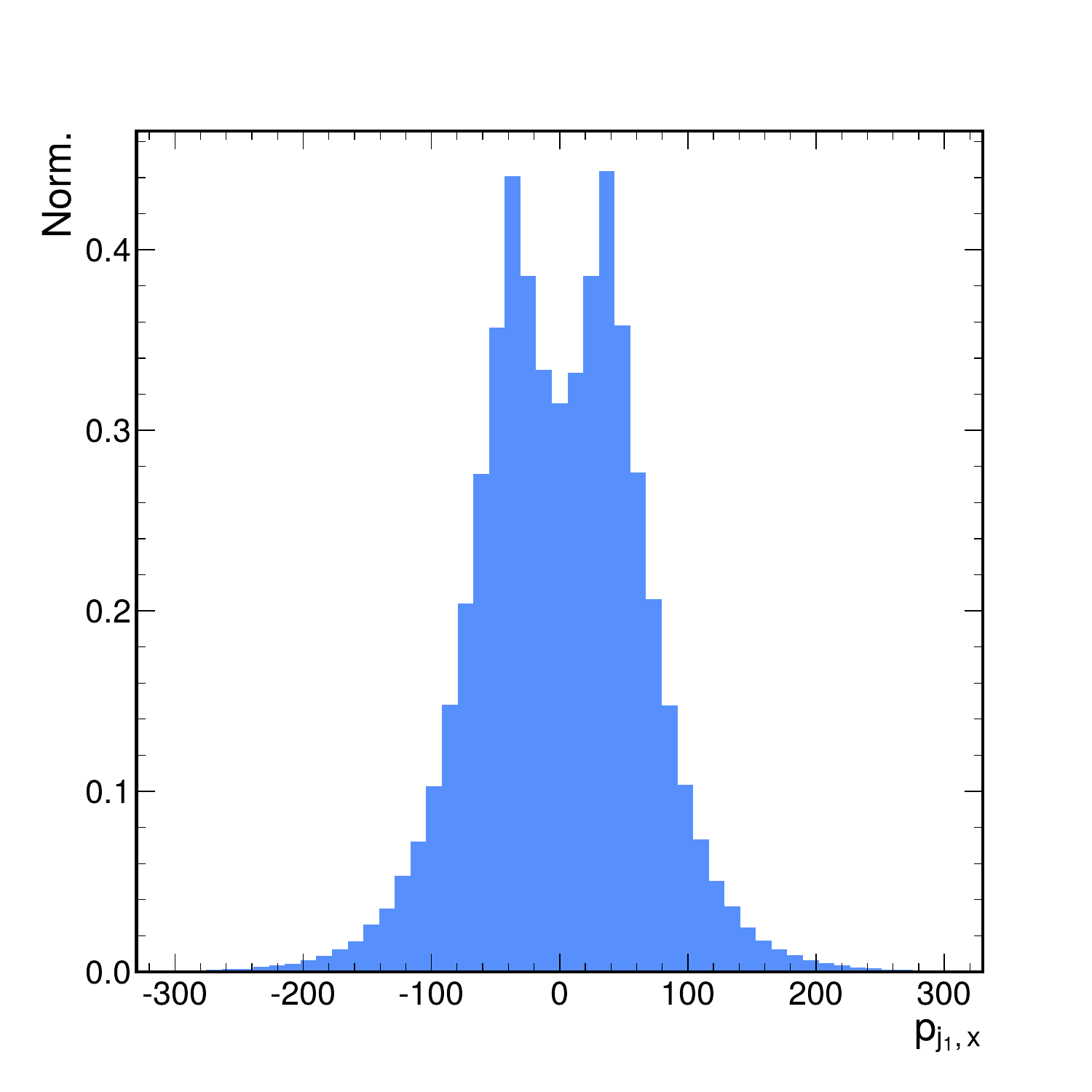}
    \includegraphics[width=0.24\textwidth]{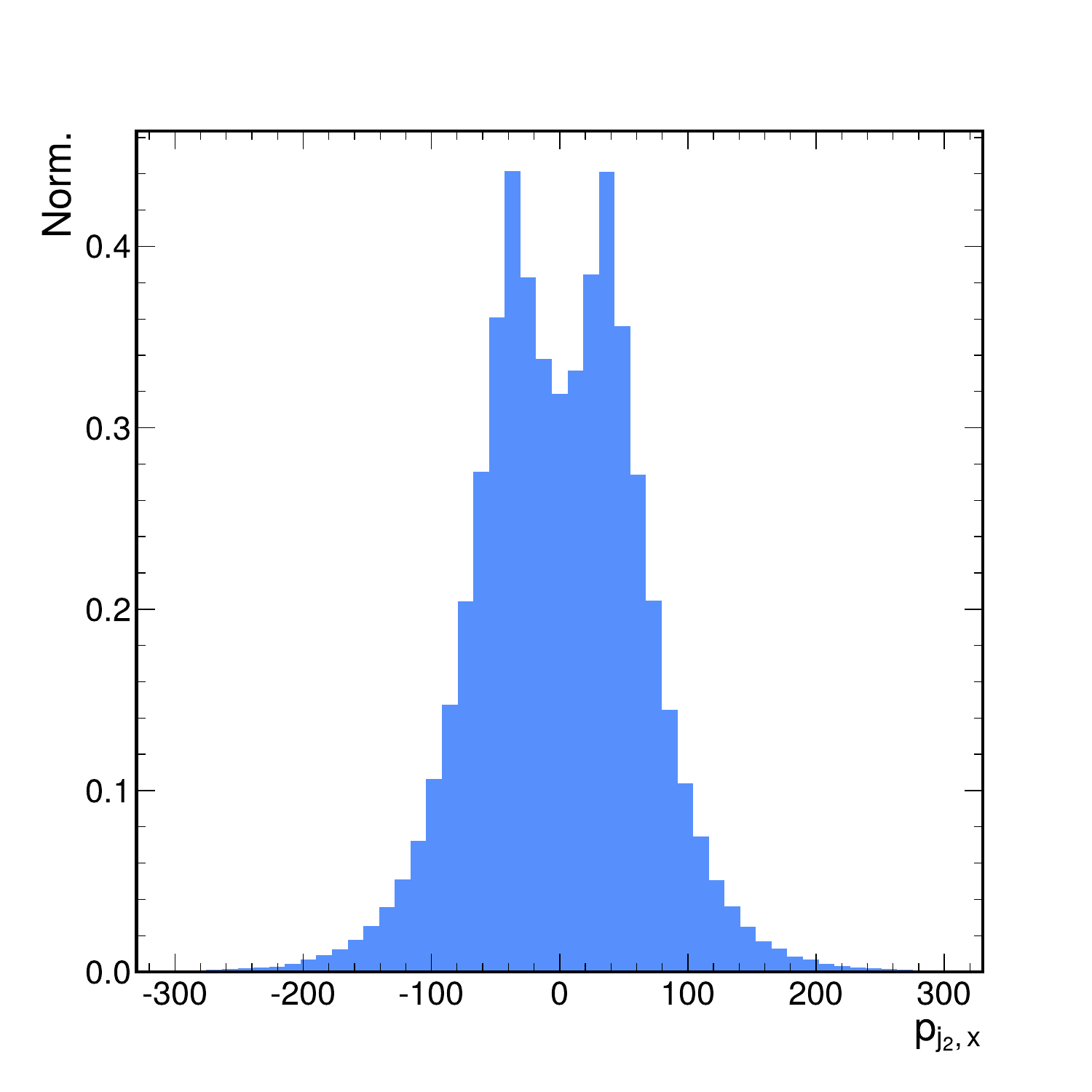}

    \includegraphics[width=0.24\textwidth]{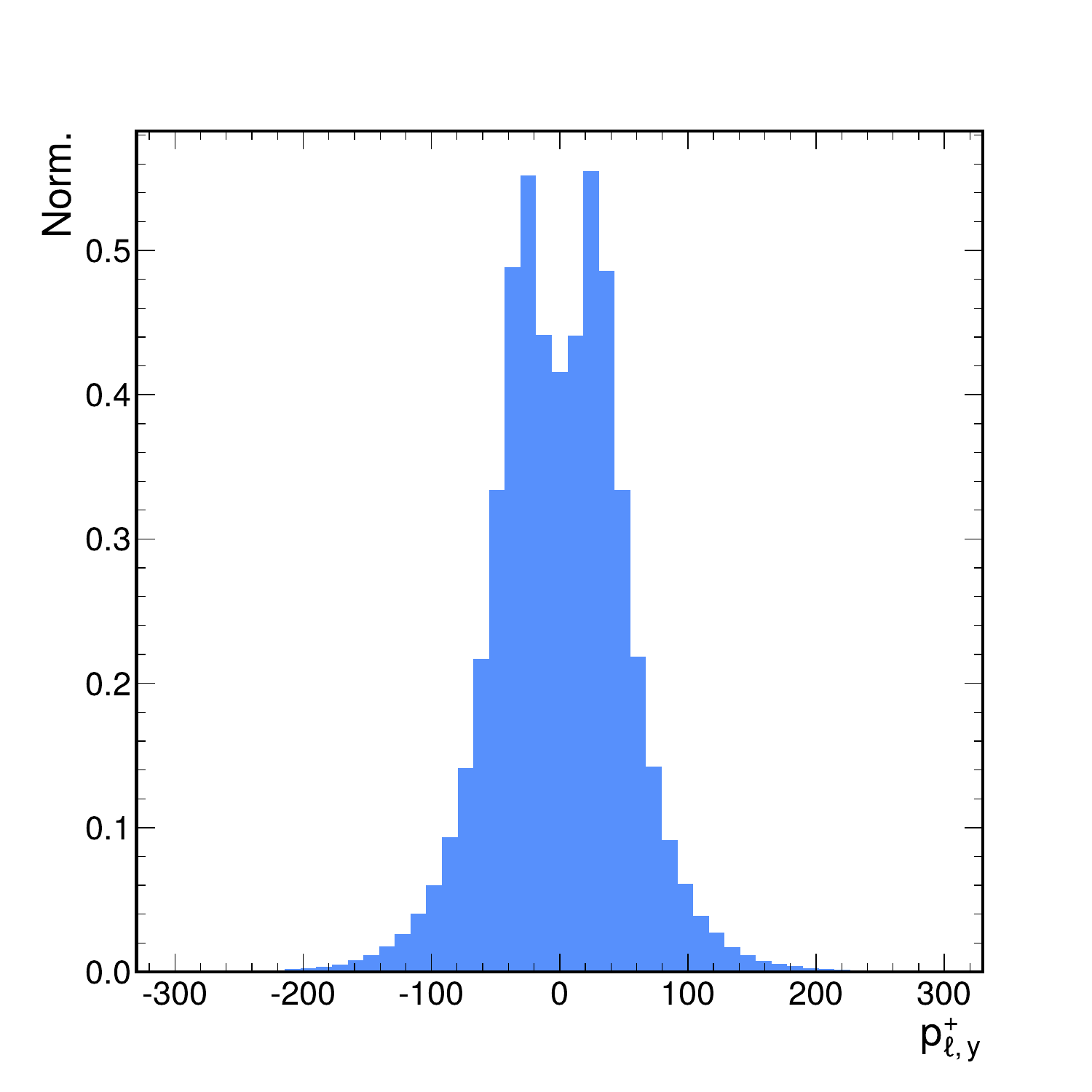}
    \includegraphics[width=0.24\textwidth]{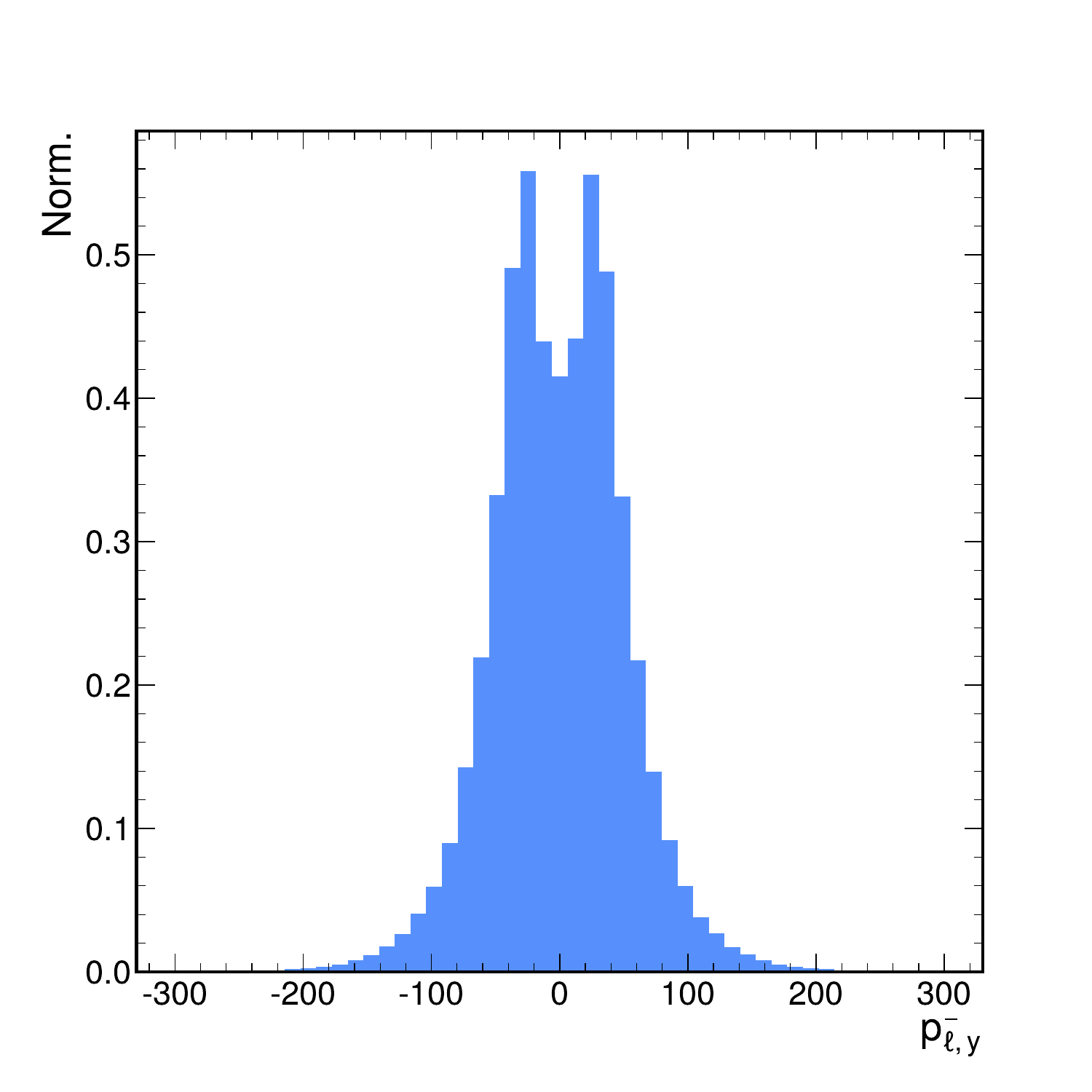}
    \includegraphics[width=0.24\textwidth]{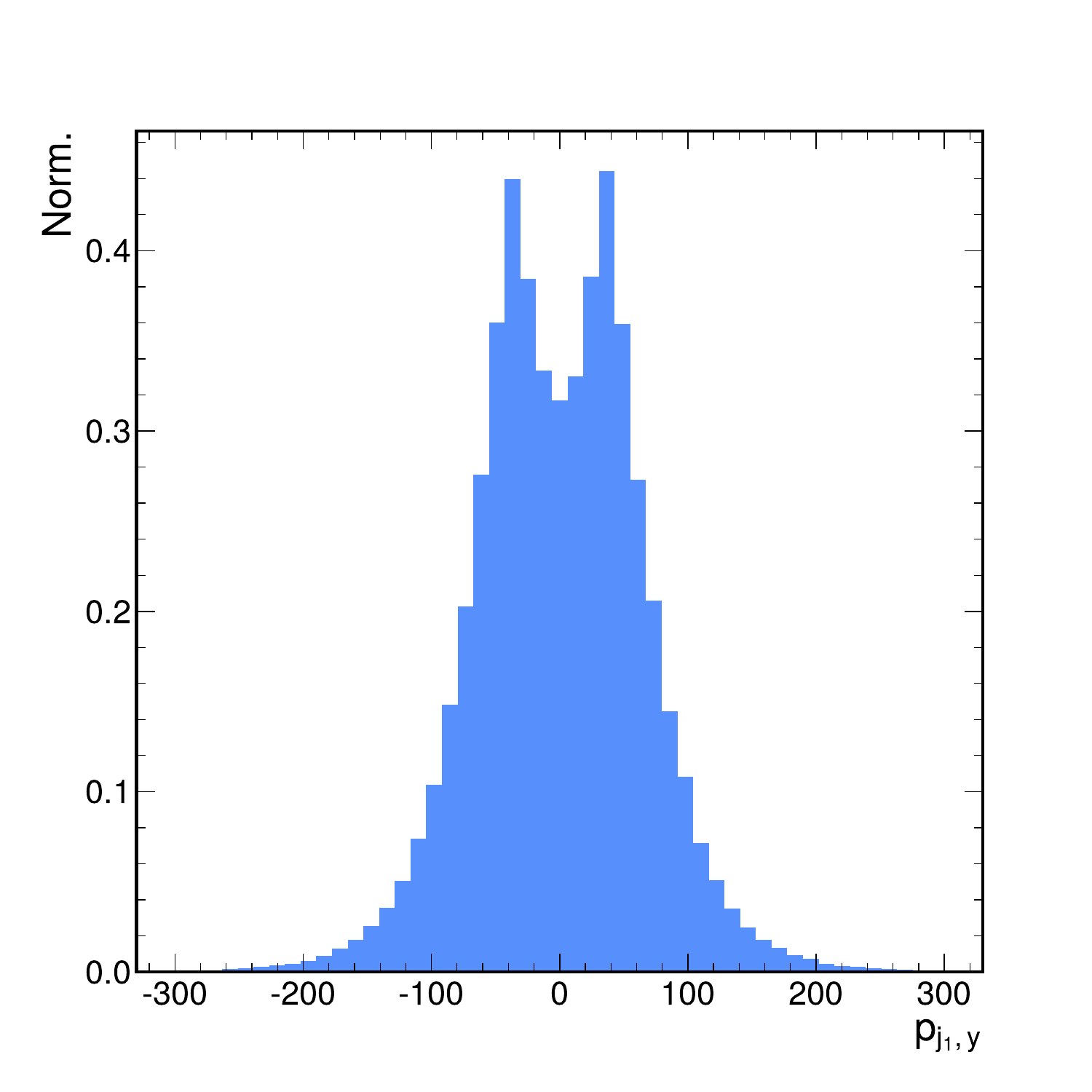}
    \includegraphics[width=0.24\textwidth]{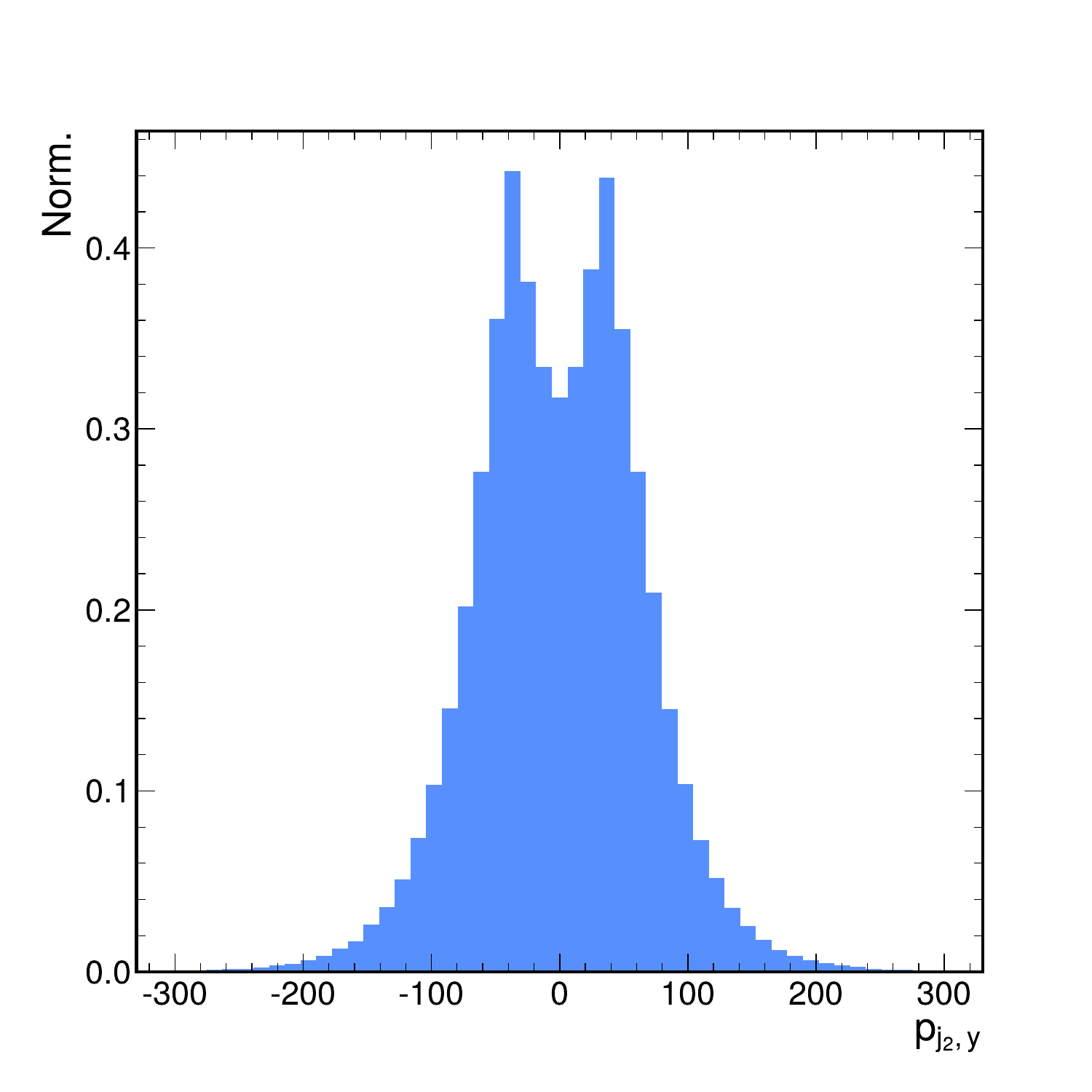}

    \includegraphics[width=0.24\textwidth]{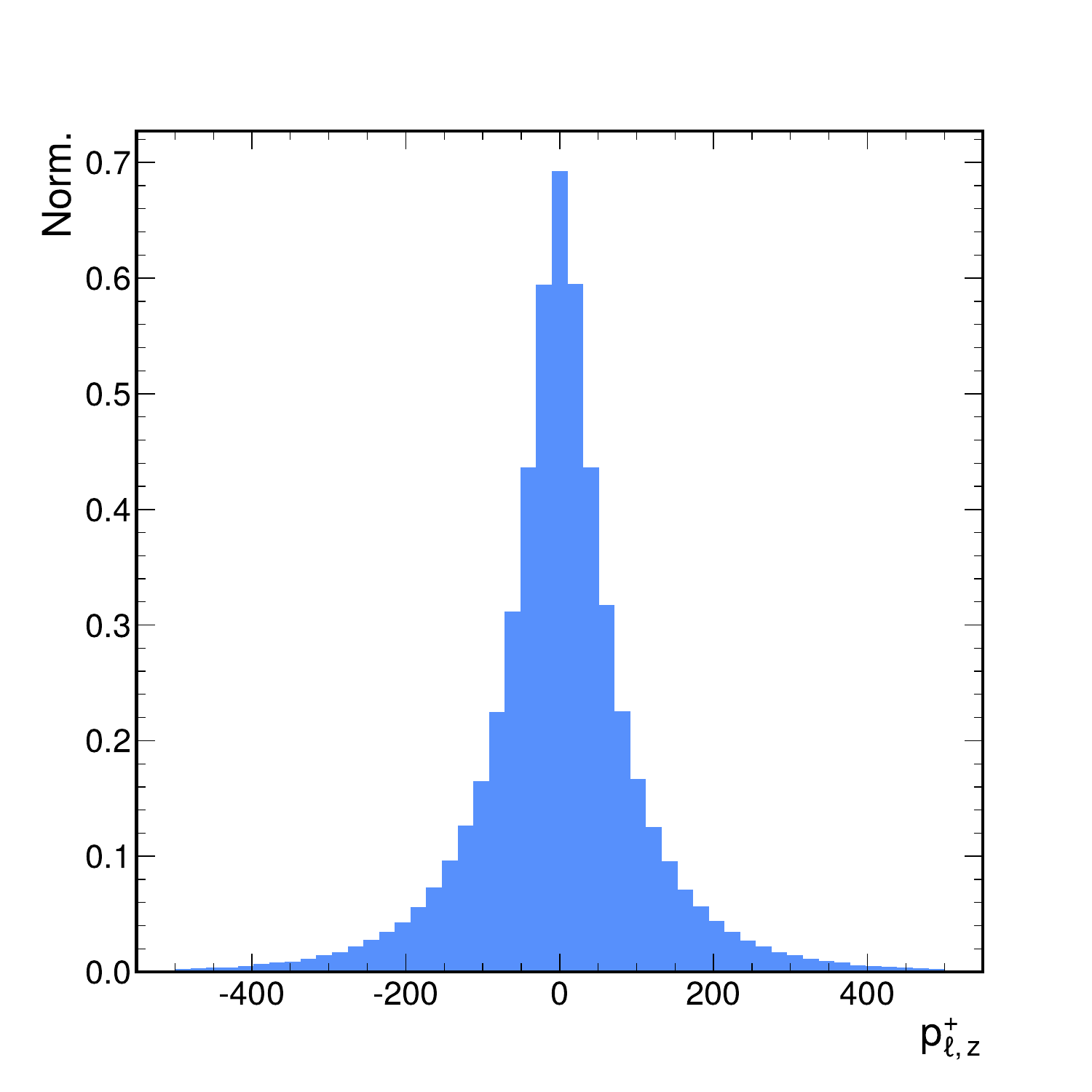}
    \includegraphics[width=0.24\textwidth]{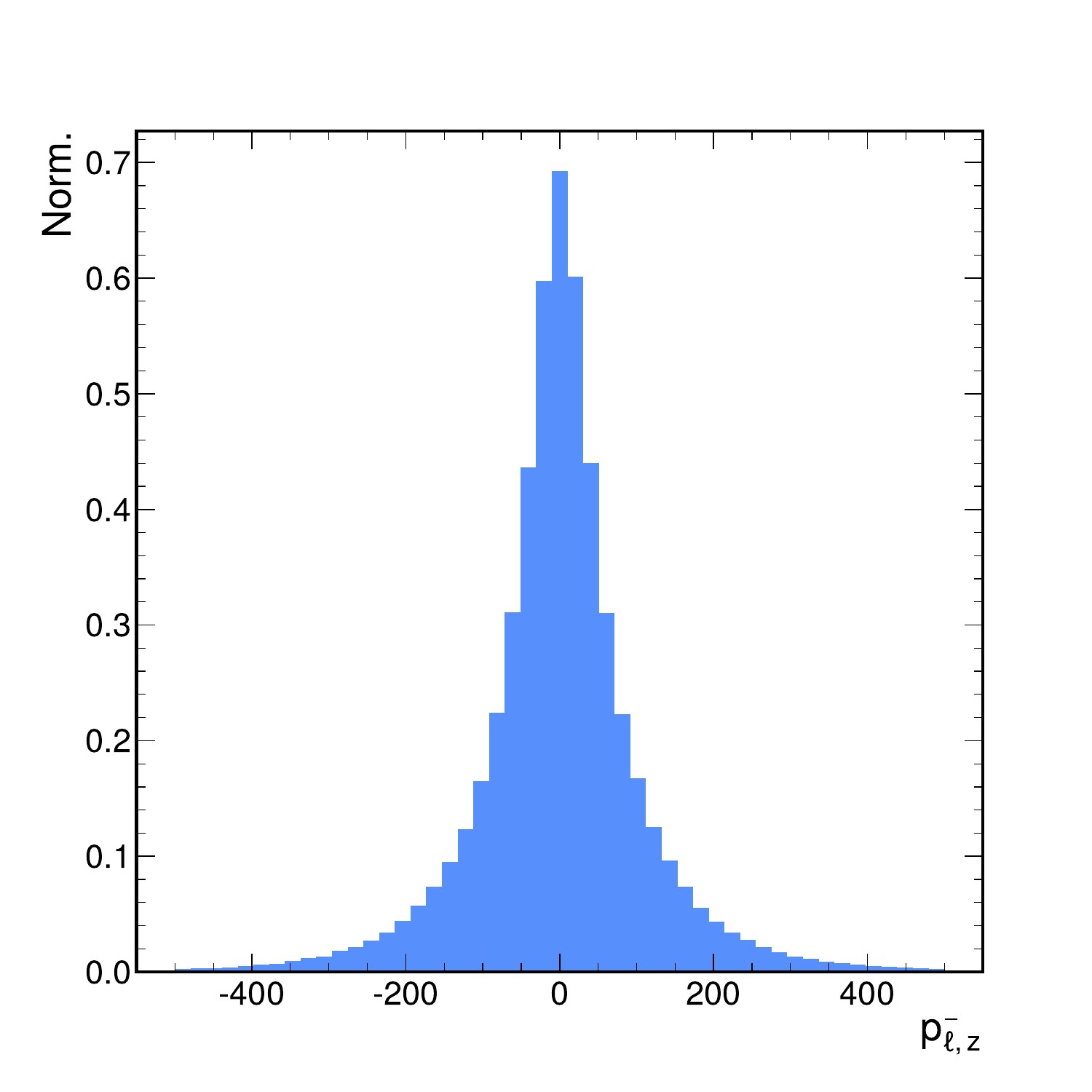}
    \includegraphics[width=0.24\textwidth]{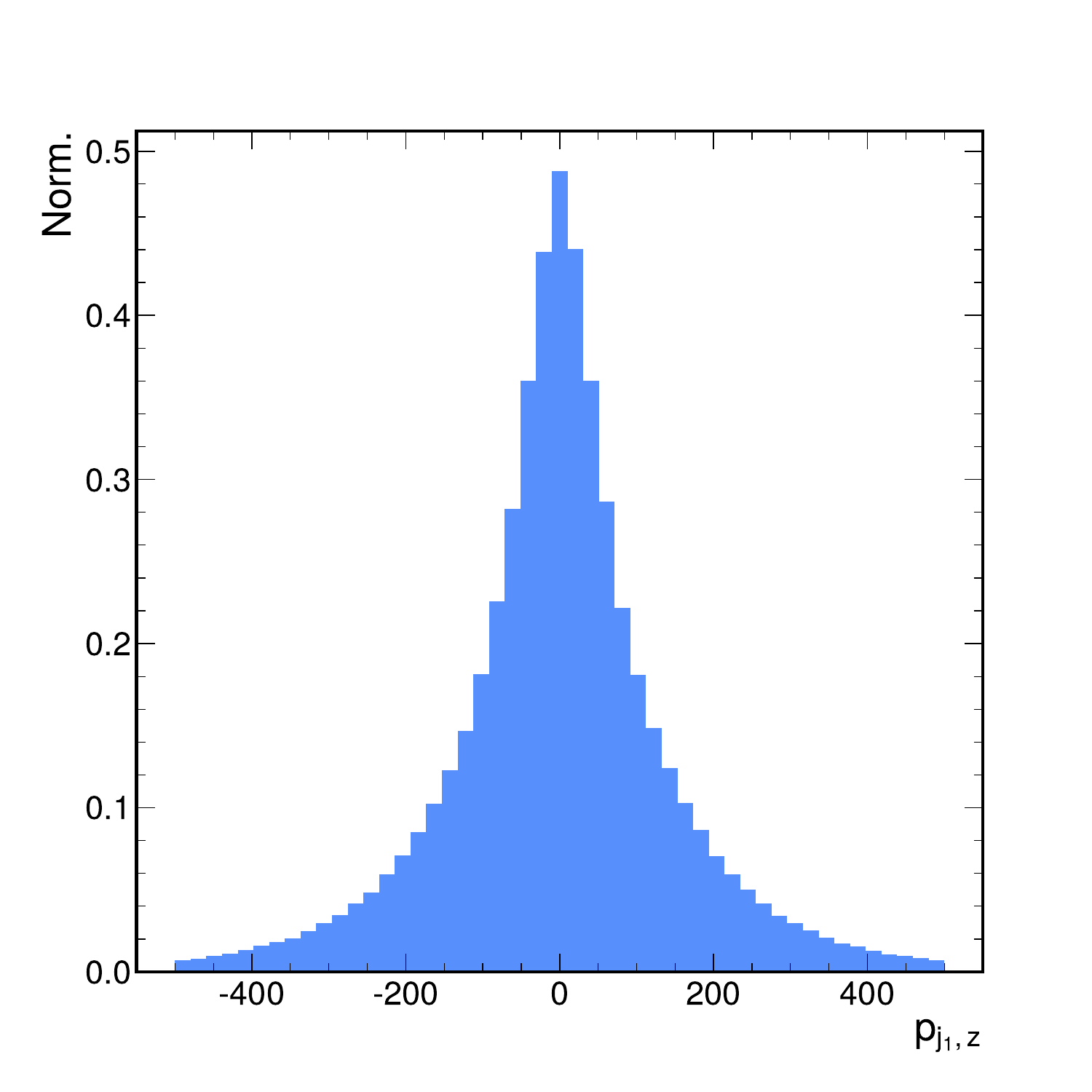}
    \includegraphics[width=0.24\textwidth]{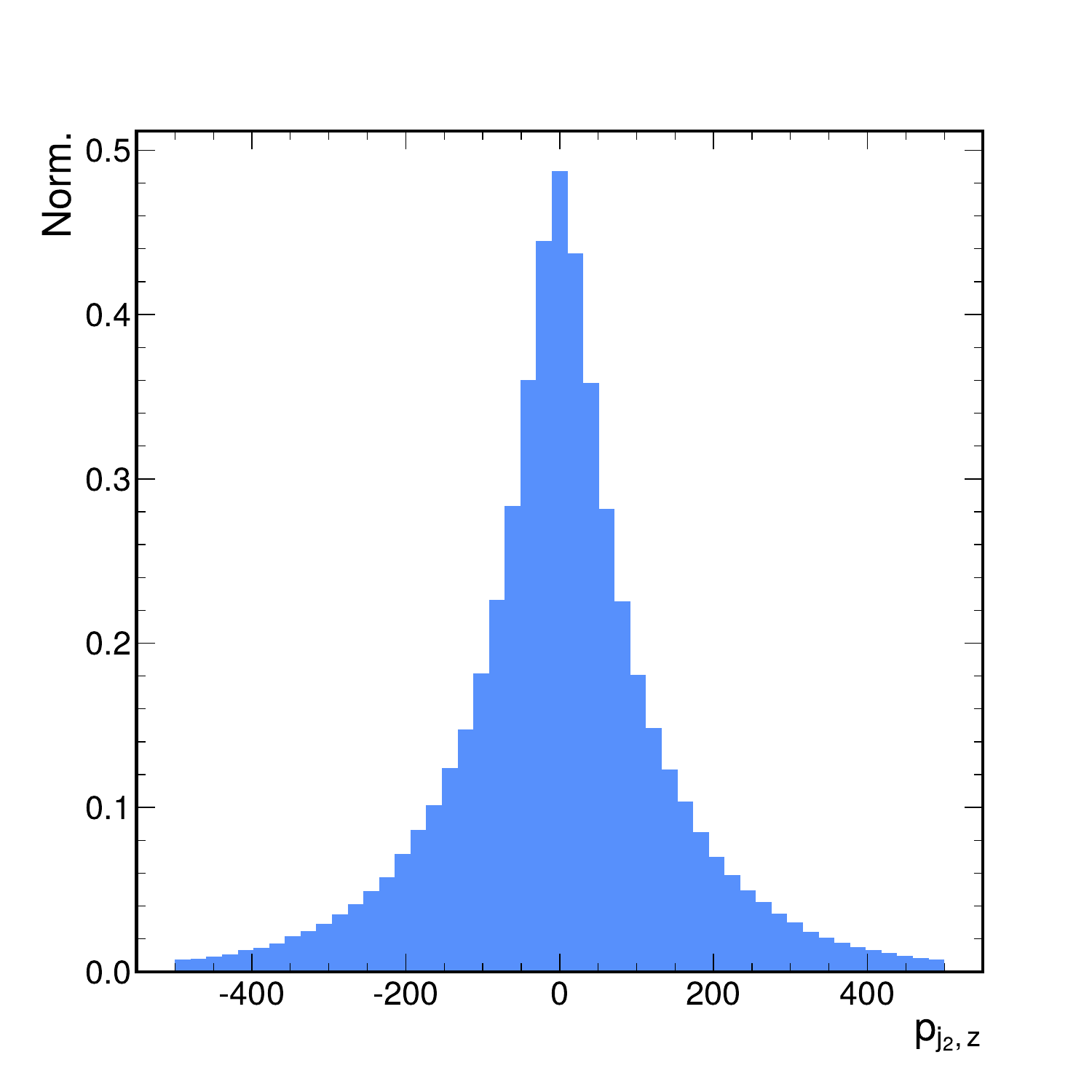}

    \includegraphics[width=0.24\textwidth]{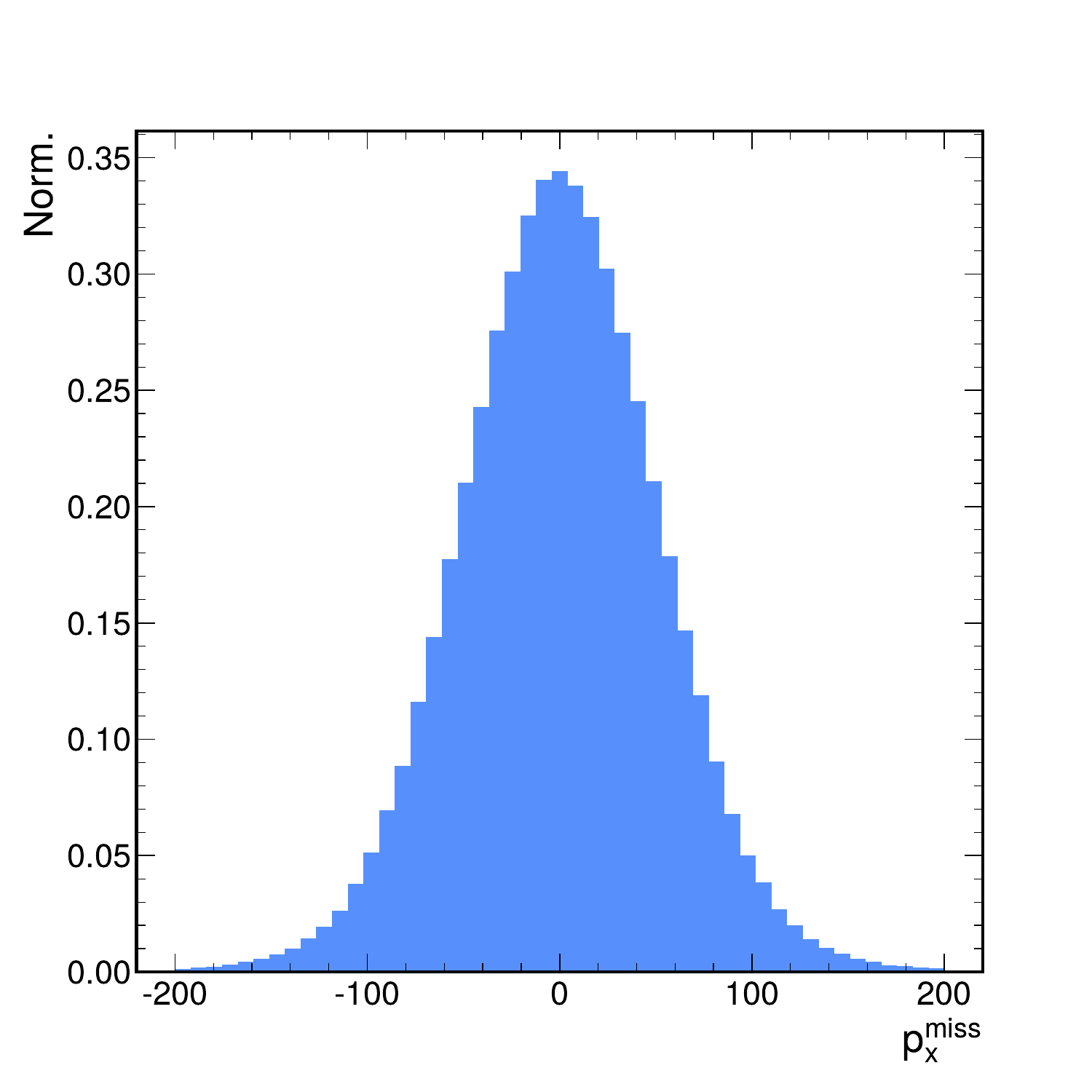}
    \includegraphics[width=0.24\textwidth]{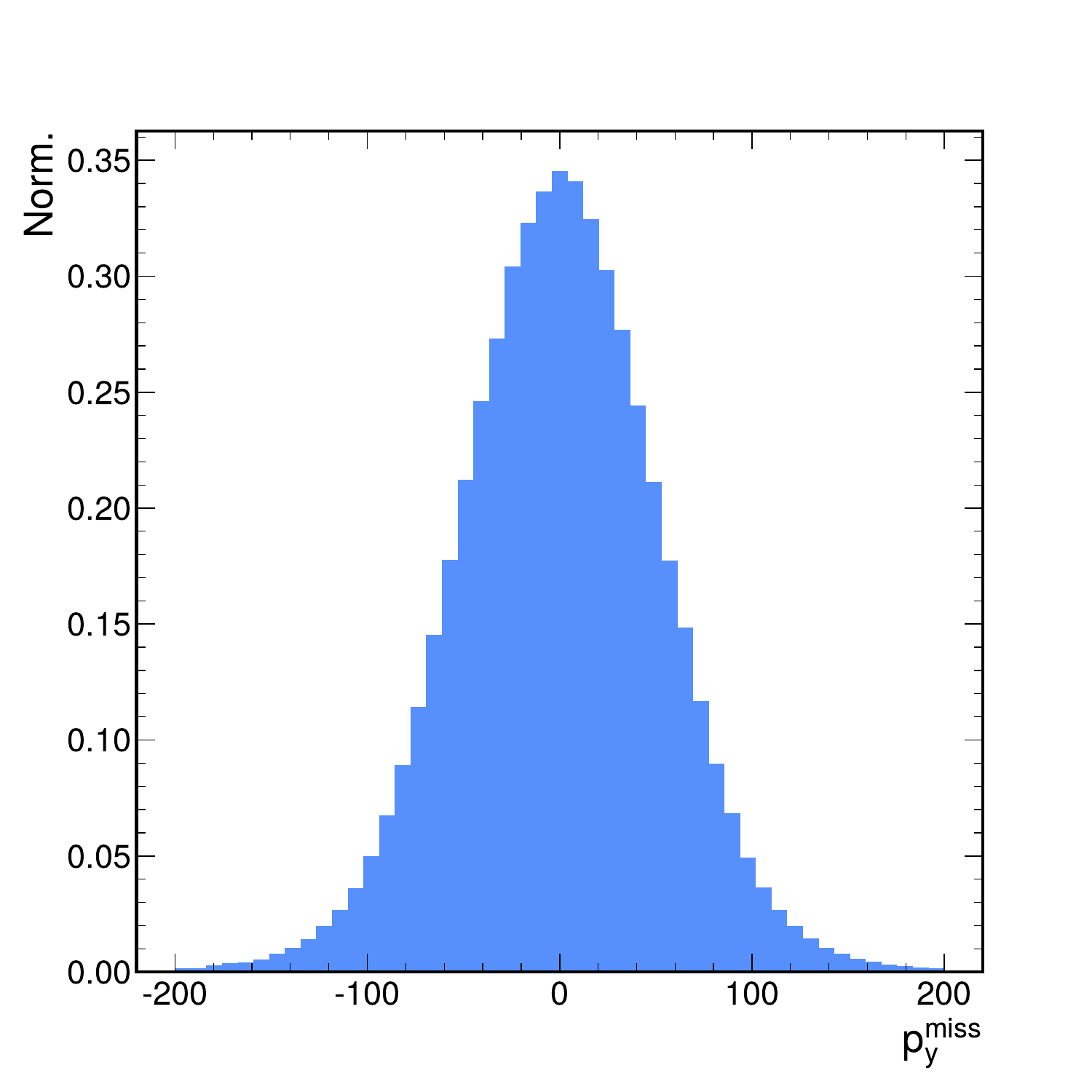}
    \caption{ Distribution of \ttbar events under the SM hypothesis of the input variables used in our algorithm. The first three rows show the $x$ (first row), $y$ (second row) and $z$  (third row) components of the three-momentum of the positive lepton (first column),  negative lepton (second column), first jet (third column) and second jet (fourth column). The last row shows the $x$ and $y$ components of $\overrightarrow{ p_{\mathrm{T}}^{\mathrm{miss}}}$. It has been checked that the linear contribution of $c_{\mathrm{tG}}^I$ does not modify the one-dimensional distribution of these variables.  }
    \label{fig:ttbar_inputvars}

\end{figure*}

\begin{figure}
    \centering
    \includegraphics[width=\linewidth]{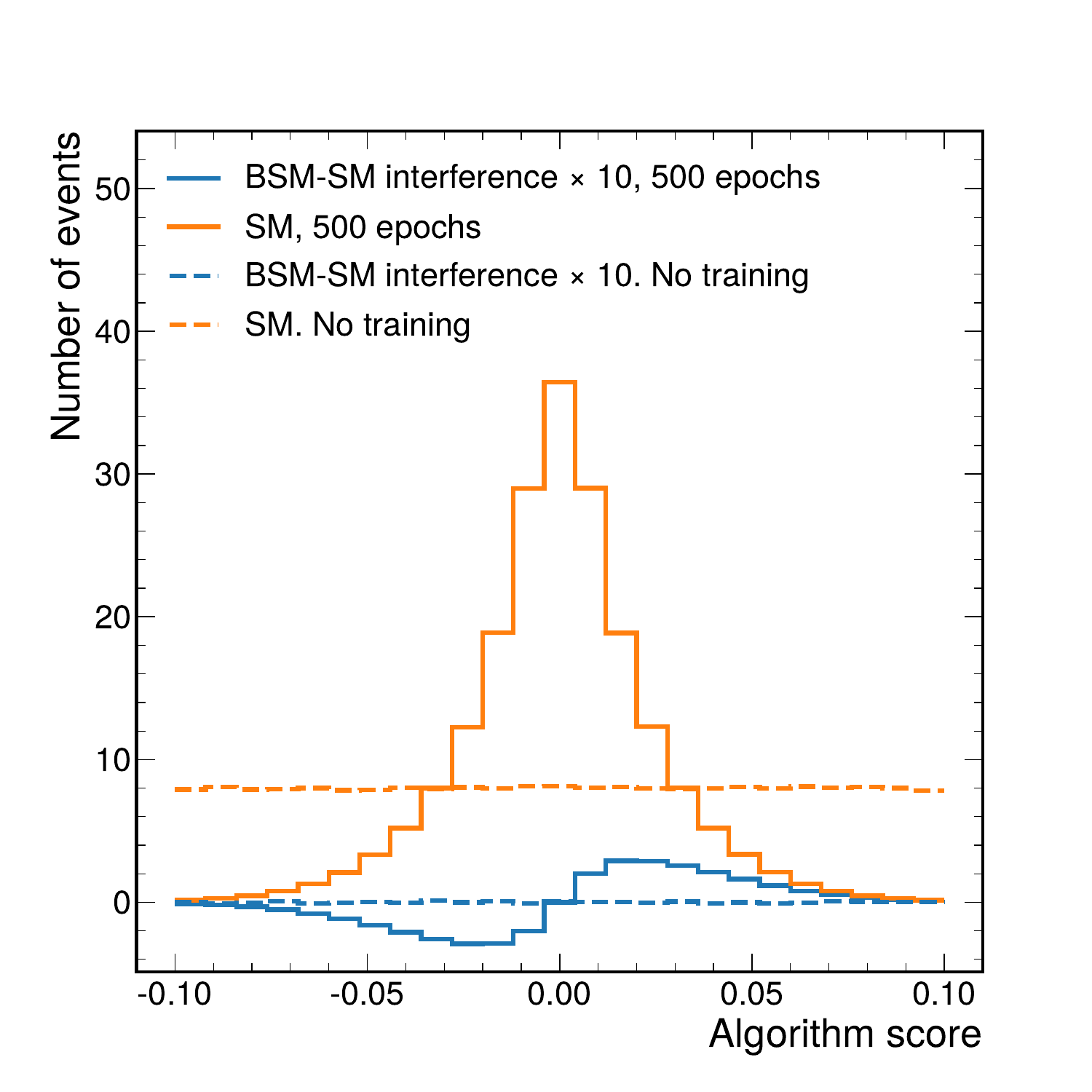}
    \caption{Distribution of the algorithm score before (dashed lines) and after (full lines) training, for simulated events distributed as the SM hypothesis, and the linear contribution when $c_{\mathrm{tG}}^I=10$.}
    \label{fig:ttbar_results_1}
\end{figure}

Our space of input variables $\mathcal{D}$ is spanned by the following variables, whose distribution is shown in Fig.~\ref{fig:ttbar_inputvars}: the three-momentum vector of the positively-charged leptons $\overrightarrow{p_{\ell}^+}$, the three-momentum vector of the negatively-charged leptons $\overrightarrow{p_{\ell}^-}$, the three-momenta of two selected jets $\overrightarrow{p_{j_1}}$ and $\overrightarrow{p_{j_2}}$, and the missing transverse momentum in the transverse plane $\overrightarrow{ p_{\mathrm{T}}^{\mathrm{miss}}}$. As the two selected jets, we take the two leading \Pb-tagged jets. If only one jet in the event is \Pb-tagged, we select it along with the leading non-\Pb-tagged jet. The indices of the two jets are assigned at random in order to have a consistent representation of the \CP transformation, $h_{\CP}$. We choose $h_{\CP}$ as:

\begin{figure*}
\includegraphics[width=0.32\textwidth]{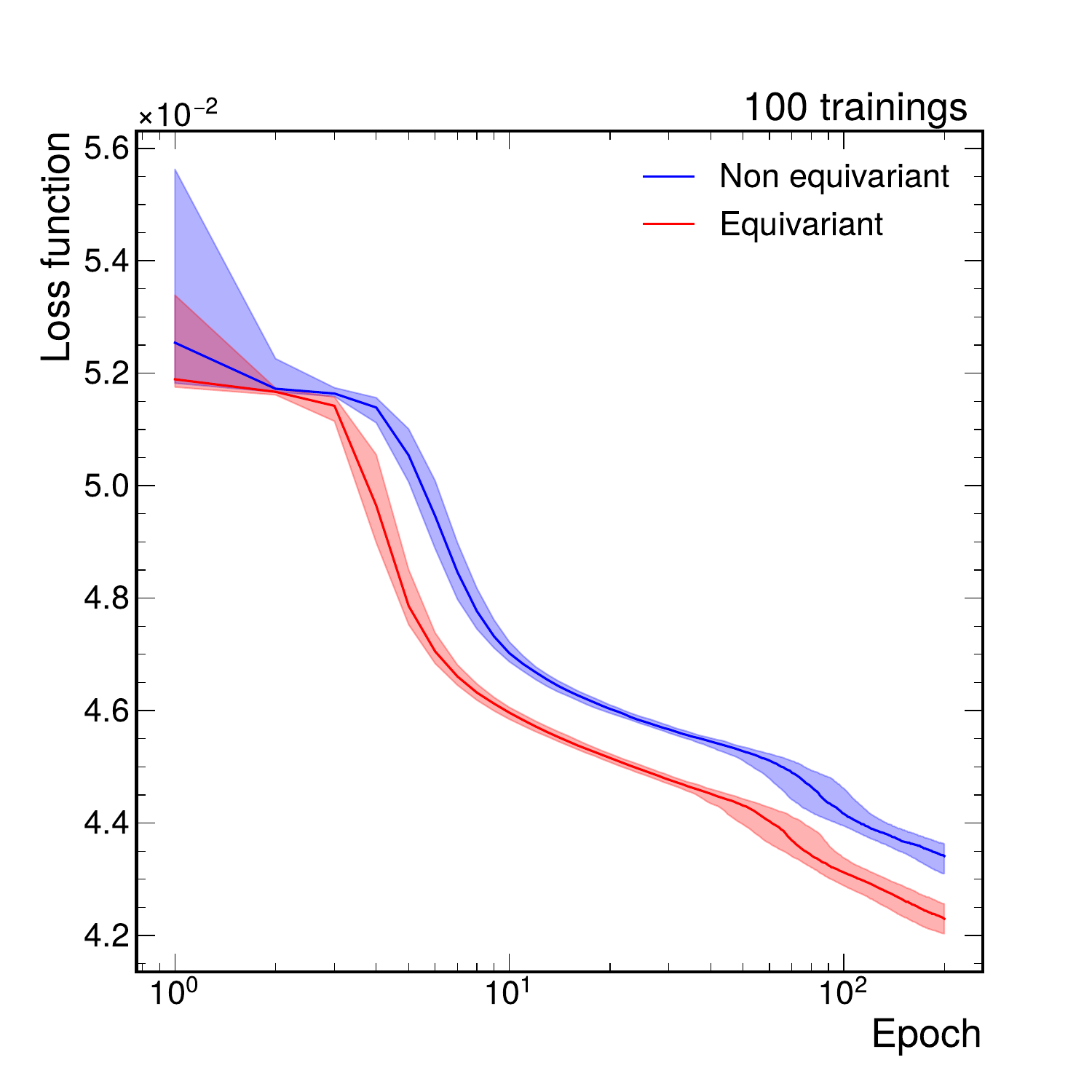}
\includegraphics[width=0.32\textwidth]{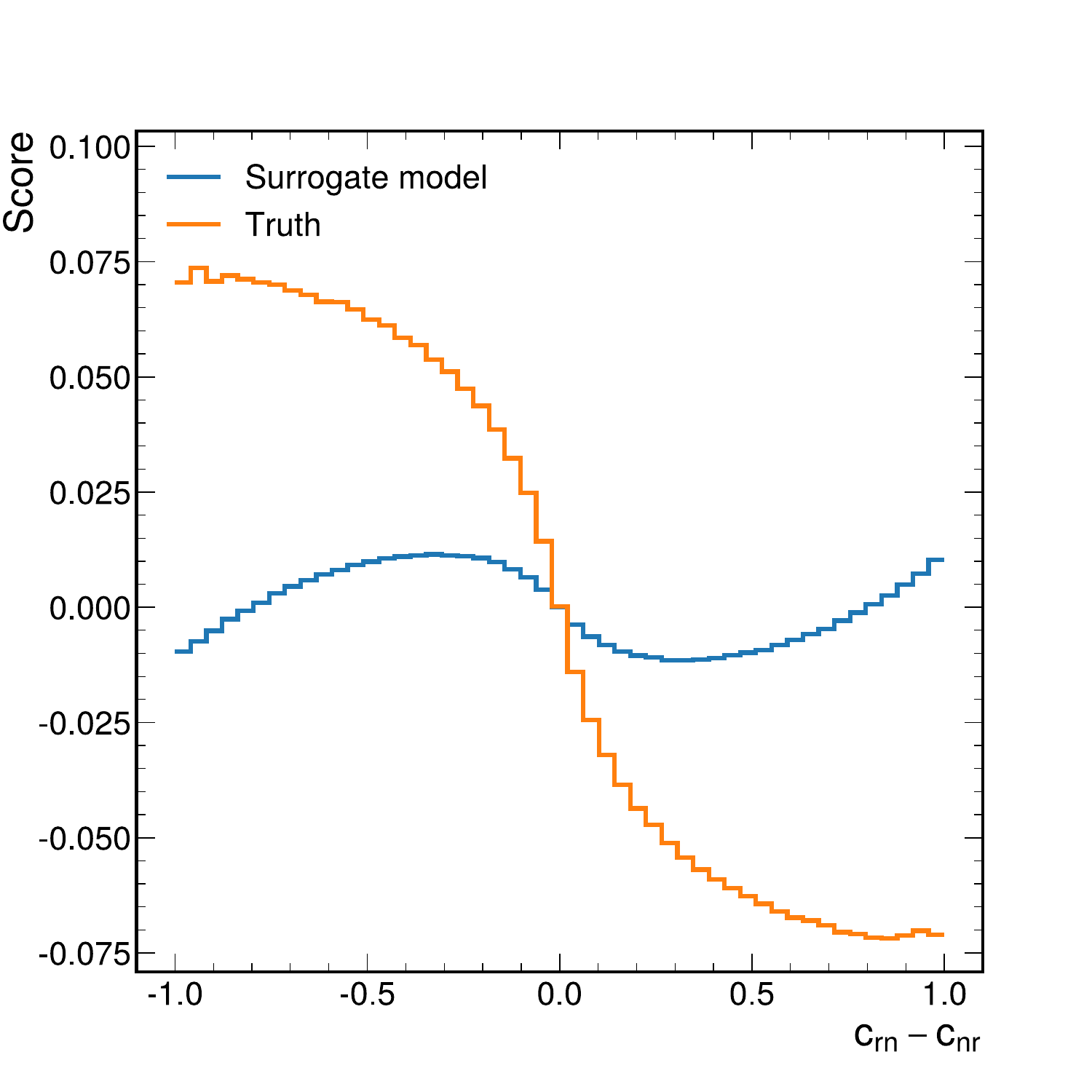}
\includegraphics[width=0.32\textwidth]{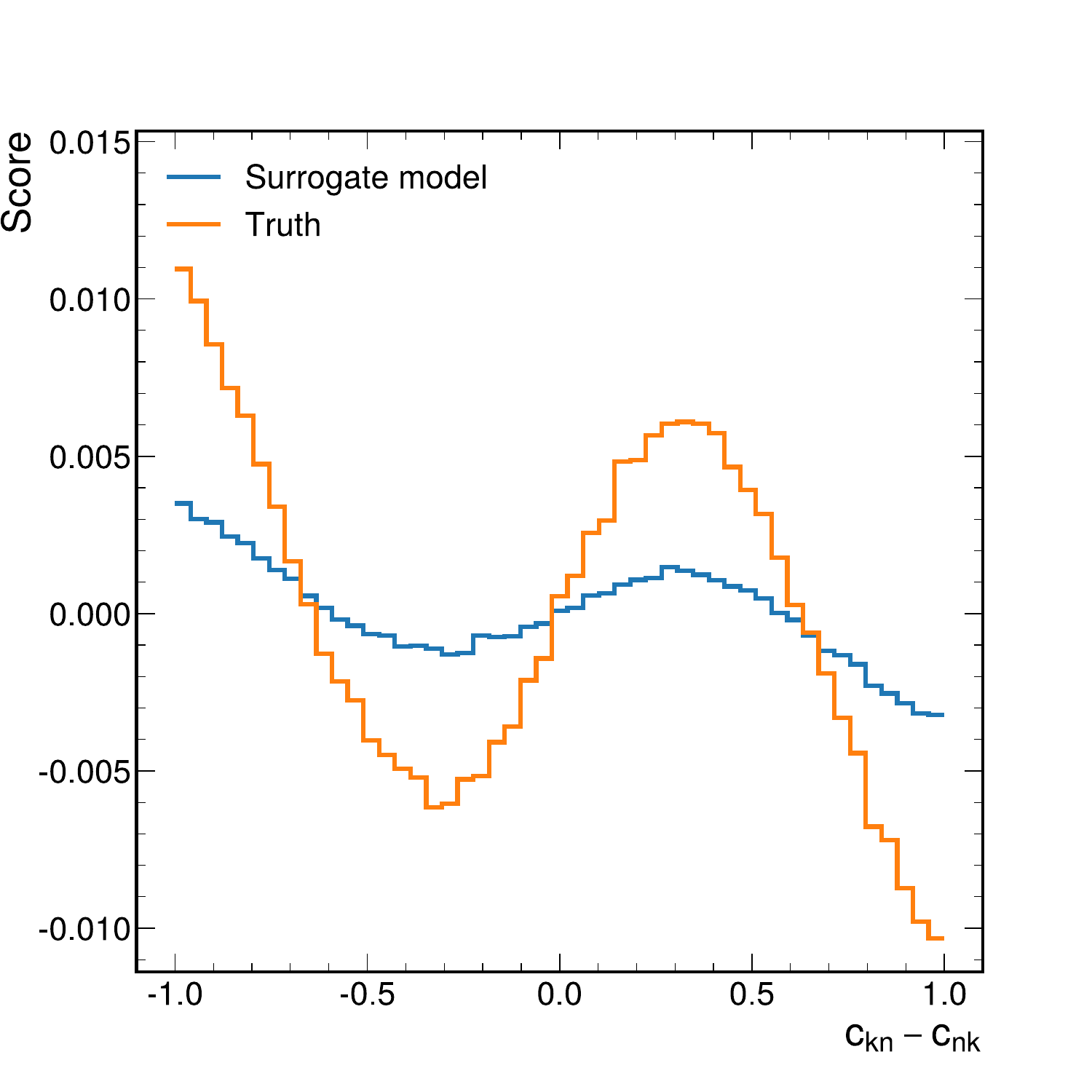}
\caption{\label{fig:ttbar_results_2} Left: Evaluation of the loss function in the test dataset as a function the number of epochs that have been trained for the equivariant and nonequivariant algorithms. The lines show the median of 100 independent trainings of each algorithm and the bands span the 0.16 and 0.84 quantiles. Center and right: score mean as a function of $\mathrm{c}_{\mathrm{rn}}-\mathrm{c}_{\mathrm{nr}}$ (center) and $\mathrm{c}_{\mathrm{kn}}-\mathrm{c}_{\mathrm{nk}}$ (right), using the true and surrogate model.}
\end{figure*}

\begin{align}
    h_{\CP}\left( \overrightarrow{p_{\ell}^+}, \overrightarrow{p_{\ell}^-}, \overrightarrow{p_{j_1}}, \overrightarrow{p_{j_2}}, \overrightarrow{ p_{\mathrm{T}}^{\mathrm{miss}}} \right) \nonumber \\
    = \left( -\overrightarrow{p_{\ell}^-}, -\overrightarrow{p_{\ell}^+}, -\overrightarrow{p_{j_2}}, -\overrightarrow{p_{j_1}}, -\overrightarrow{ p_{\mathrm{T}}^{\mathrm{miss}}} \right)\,.
\end{align}

We train our algorithm by minimizing the loss function in equation~\eqref{eq:loss}, associated to the linear contribution of $c_{\mathrm{tG}}^I$. The distribution of the regressed score is shown in figure~\ref{fig:ttbar_results_1} for events distributed under the SM hypothesis and the interference contribution. As described earlier, the equivariance condition imposes that the SM distribution as a function of the score is an even function, while the interference is an odd function, which is shown in the figure. In addition, SM-like events tend to have values closer to zero, while events resembling the \CP-odd contribution tend to take larger absolute values. 

To illustrate one of the properties of the algorithm, we also depict the score of the algorithm before any training has been performed, with all parameters of the $g$ function set to their initial random values. In that case, the equivariance property still holds and the SM (linear) contributions are distributed as an odd (even) function. While this example is of little practical relevance, it serves us to illustrate that the algorithm will produce \CP-odd observables, regardless of the convergence of the training, a crucial property in practice, as discussed in the previous section.

We also show that imposing equivariance as an inductive bias in the algorithm improves the numerical convergence in the training. To do so, we train 100 instances of our equivariant algorithm and another 100 of a nonequivariant algorithm. The nonequivariant algorithm minimizes the same loss function and the same function $g$ as the equivariant case. The initial weights and biases of $g$ are randomly chosen in each instance of the trainings, while all the hyperparameters take the same values. Figure~\ref{fig:ttbar_results_2} shows the median of the loss function evaluated in the test dataset as a function of the training epoch, along with bands containing 68\% of the trainings. The results show that the training converges significantly faster when considering the equivariant algorithm: the median of the nonequivariant algorithm takes between 40\% and 300\% fewer epochs than the equivariant algorithm to reach a given loss function value.

Finally, in order to interpret the features that the algorithm is learning, we check the model against observables specifically designed to capture \CP violation effects in \ttbar production. We consider $\mathrm{c}_{\mathrm{rn}}-\mathrm{c}_{\mathrm{nr}}$  and $\mathrm{c}_{\mathrm{kn}}-\mathrm{c}_{\mathrm{nk}}$, which are described in Ref.~\cite{Bernreuther:2015yna}. These observables are related to angles of the leptons in an specific reference frame, rely on the reconstruction of the \ttbar system to be constructed, and are sensitive to the linear contribution of $c_{\mathrm{tG}}^I$.

Figure~\ref{fig:ttbar_results_2} shows the average score as a function of each of these variables for the true model and for the surrogate model learned by our algorithm. These magnitudes quantify, for each of the models, the dependence of the distribution of the different variables as a function of the WCs for infinitely small values of the WCs, which is given by the linear contributions only. Obtaining the same value for two models means that the surrogate model is able to learn the dependence introduced by the linear contribution that is predicted by the simulation. The algorithm is able to partially learn the trend predicted by the true model, especially for $\mathrm{c}_{\mathrm{kn}}-\mathrm{c}_{\mathrm{nk}}$, although not fully. We have verified that this is not due to limitations introduced by our algorithm, by training a similar model that uses all parton-level information as an input and that is able to fully learn the behavior predicted by the true model. In addition, we have checked that the nonequivariant model learns a similar behavior as the equivariant one, but needs more iterations to be trained, as discussed in the previous paragraph. 

\begin{figure*}
\includegraphics[width=0.32\textwidth]{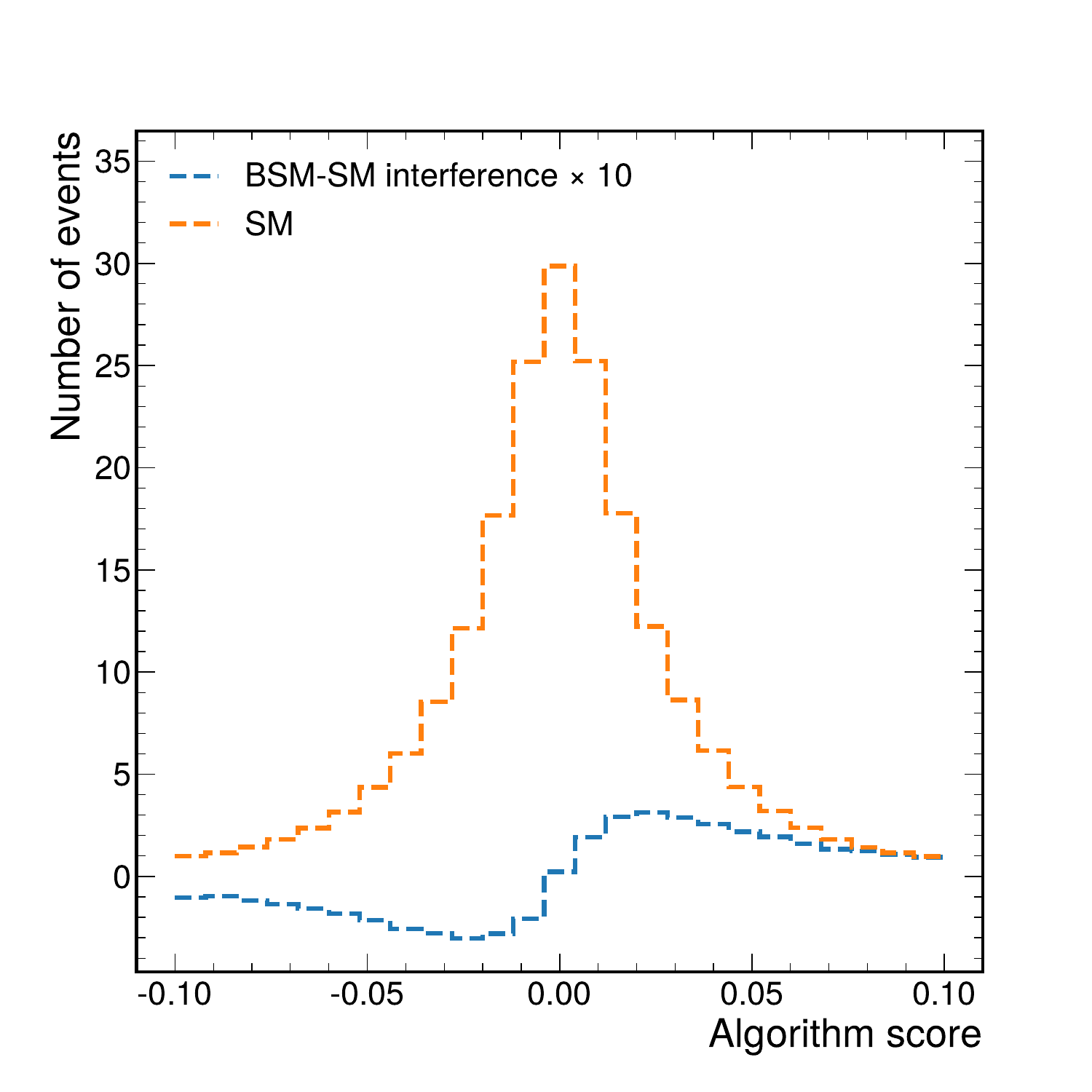}
\includegraphics[width=0.32\textwidth]{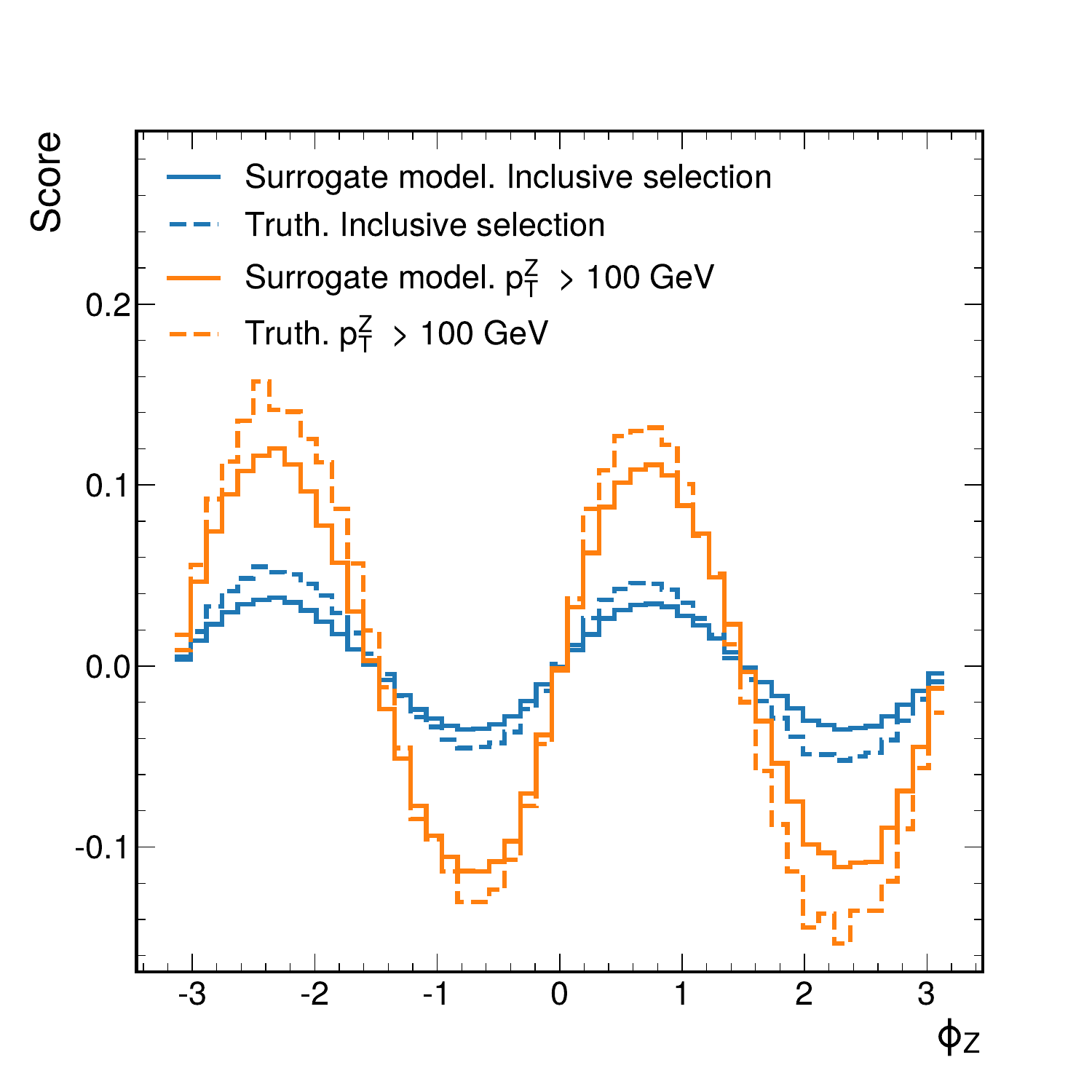}
\includegraphics[width=0.32\textwidth]{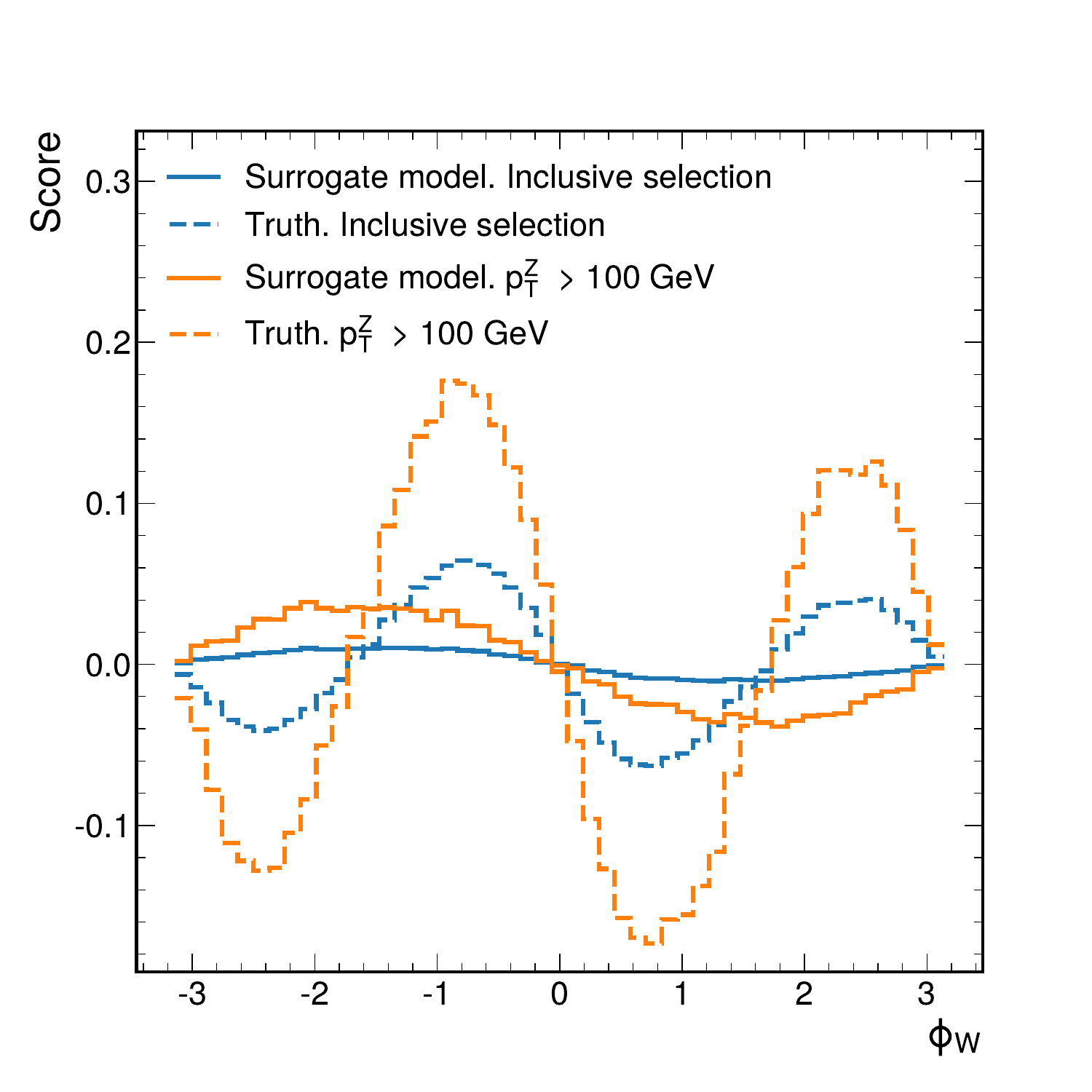}
\caption{\label{fig:wz_results} Left: Distribution of the algorithm score for simulated events distributed as the SM hypothesis, and the linear contribution when $c_{\tilde{\mathrm{W}}}=10$. Center and right: score mean as a function of $\phi_{\mathrm{Z}}$ (left) and $\phi_{\mathrm{W}}$ (right), using the true and surrogate model.}
\end{figure*}

We conclude that the fact that our algorithm does not fully learn the true model is due to the impossibility of unambiguously reconstructing the \ttbar system in the dileptonic decay mode, stemming from the presence of two undetected neutrinos, the assignment of the selected jets to each of the top quarks, and the presence of additional jets produced in the  initial- and final-state radiation. To a lesser extent, the power of the algorithm is also bounded by the limited expressivity of the chosen functional form for $g$. Several advanced algorithms have been developed to improve the reconstruction of the \ttbar system~\cite{Sonnenschein:2006ud,CMS:2019esx,CMS:2022emx} that could be used as inputs to our algorithm, but we consider this beyond the scope of this work.

\subsection{CP violation in \texorpdfstring{$\WZ$}{WZ} production} 
\label{sec:wz} 

We consider now \WZ production in its final state with three leptons, and the effect of the \CP-odd operator $\mathcal{O}_{\tilde{\mathrm{W}}} = \epsilon_{ijk}  \tilde{W}_{\mu}^{i\nu} W_{\nu}^{j\rho}W_{\rho}^{k\mu}$. Similar new physics scenarios have been studied in the framework of anomalous couplings in Ref.~\cite{CMS:2021icx}, but \CP-odd observables were not studied. Reference~\cite{Panico:2017frx} describes some \CP-odd observables specifically for this process, which rely on the angles between each of the bosons in the decay plane and the beam plane. We use our algorithm to extend the reach of the searches profiting from those angular variables, as well as the energy growth expected in these operators. 

We selected events with three leptons with $\pt > 15$ GeV and $|\eta| < 2.5$, and we require the leading lepton to have $\pt > 25$ GeV. Out of the three selected leptons, we define a $\PZ$ boson candidate by picking the two opposite-sign same-flavor leptons with their invariant mass closer to the $\PZ$ boson mass. As input variables to our algorithm, we take the three-momentum of the positive and negative lepton of the $\PZ$ candidate, $\overrightarrow{p_{\ell^+}^{\PZ}}$ and $\overrightarrow{p_{\ell^-}^{\PZ}}$; the three-momentum of the third lepton $\overrightarrow{p_{\ell}^{\PW}}$ and its charge $Q^{\PW}$, and the missing momentum vector in the transverse plane $\overrightarrow{ p_{\mathrm{T}}^{\mathrm{miss}}}$. We choose $h_{\CP}$ as:

\begin{align}
    h_{\CP}\left( \overrightarrow{p_{\ell^+}^{\PZ}}, \overrightarrow{p_{\ell^-}^{\PZ}}, \overrightarrow{p_{\ell}^{\PW}}, Q^{\PW}, \overrightarrow{ p_{\mathrm{T}}^{\mathrm{miss}}} \right) = \nonumber \\  \left(  -\overrightarrow{p_{\ell^-}^{\PZ}}, -\overrightarrow{p_{\ell^+}^{\PZ}}, -\overrightarrow{p_{\ell}^{\PW}}, -Q^{\PW}, -\overrightarrow{ p_{\mathrm{T}}^{\mathrm{miss}}} \right)\,.
\end{align}

We show the score of our algorithm after being trained in Fig.~\ref{fig:wz_results}, resulting in a good separation power between the SM hypothesis and the linear contribution. In addition, the distribution of events under the SM hypothesis has an even distribution as a function of this score, while the linear contribution has an odd distribution, as expected from the equivariance property of the algorithm. 

We contrast our results against known observables sensitive to the linear term, described in Ref.~\cite{Panico:2017frx}. The linear contribution introduced by the $\mathcal{O}_{\tilde{\mathrm{W}}}$ induces a modulation in the $\phi_{\mathrm{Z(W)}}$ angle, defined as the angle between the \PZ (\PW) boson decay plane and the beam plane, in the \WZ rest frame. In order to check whether our algorithm is learning this modulation, we show in figure~\ref{fig:wz_results} the average score as a function of these variables for the true model and for the surrogate model learned by our algorithm. We observe that the algorithm captures almost perfectly the modulation in $\phi_{\mathrm{Z}}$, while it does not fully capture entirely the modulation in $\phi_{\mathrm{W}}$. This behavior is, in fact, expected and described already in Ref.~\cite{Panico:2017frx} and stems from the fact that in the fully leptonic channel the decay products of the $\PW$ boson cannot be unambiguously reconstructed because of the presence an undetected neutrino. The modulation on $\phi_{\mathrm{W}}$ can be studied in events where the $\PW$ boson decays hadronically, as described in Refs.~\cite{Panico:2017frx,Chatterjee:2024pbp}.

Figure~\ref{fig:wz_results} also illustrates one of the ways our algorithm extends the sensitivity reach on top of the observables proposed in Ref.~\cite{Panico:2017frx}. In the figure, we also show the score as a function of the $\phi_{\mathrm{Z(W)}}$ observable for events passing our selection and for the subset of those events in which the $p_{\mathrm{T}}$ of the reconstructed Z boson is larger than 100 GeV. In the latter case, the amplitude of the $\phi_{\mathrm{Z(W)}}$ oscillation is larger, due to the energy growth of the linear contribution. The figure demonstrates that our algorithm is able to capture this energy growth, learning the amplitude of the $\phi_{\mathrm{Z}}$ modulation, conditionally to the kinematics of each events. 

\begin{figure}
    \centering
    \includegraphics[width=\linewidth]{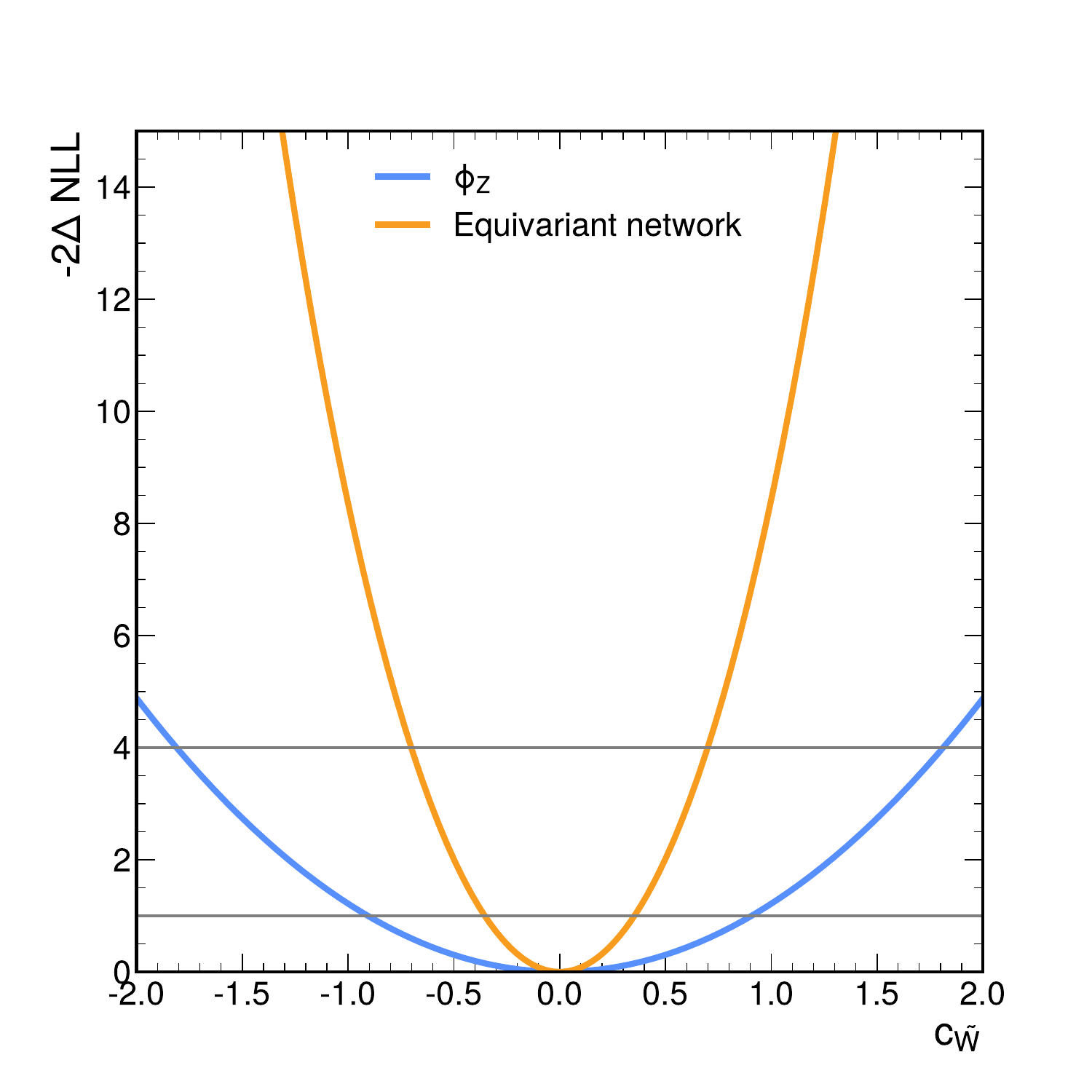}
    \caption{ Negative log-likelihood (-2$\Delta$NLL) as a function of the $c_{\tilde{\mathrm{W}}}$ parameter for an analysis based on $\phi_{\mathrm{Z}}$ (blue) and an analysis based on the score of the equivariant network (orange). The horizontal gray lines represent values where this quantity is 1 and 4. }
    \label{fig:exclusion_wz}
\end{figure}

This results in an improved sensitivity with respect to the usage of $\phi_{\mathrm{Z}}$ alone. To illustrate this, we perform two counting analyses binned in $\phi_{\mathrm{Z}}$  and in the score of our algorithm, respectively. We then build a likelihood model given by

\begin{equation}
\label{eq:likelihood_cw}
    \mathcal{L} = \prod_{j \in \mathrm{bins}} \mathcal{P}(n_j | \mu_j(c_{\tilde{\mathrm{W}}}))\,,
\end{equation}

where $\mathcal{P}$ is the Poissonian probability function with a mean $\mu_j(c_{\tilde{\mathrm{W}}})$, which denotes the number of expected events in a given bin as a function of $c_{\tilde{\mathrm{W}}}$. While this model clearly gives optimistic results, as it ignores the presence of systematic uncertainties, we note that, for strategies based on \CP-odd observables, the effect of systematic uncertainties to the linear contribution should be highly suppressed by the fact that most such uncertainties are going to have \CP-symmetric effects. In Fig.~\ref{fig:exclusion_wz}, we show the likelihood as a function of $c_{\tilde{\mathrm{W}}}$, assuming the observation of the number of events predicted by the SM. The $c_{\tilde{\mathrm{W}}}$ values for which the $2\Delta$NLL is equal to 1 and 4 can be interpreted as 68\% and 95\% confidence level intervals in those parameters, respectively, showing that our proposed method would improve the sensitivity to $c_{\tilde{\mathrm{W}}}$ by roughly a factor of three.

 \begin{figure*}
\includegraphics[width=0.32\textwidth]{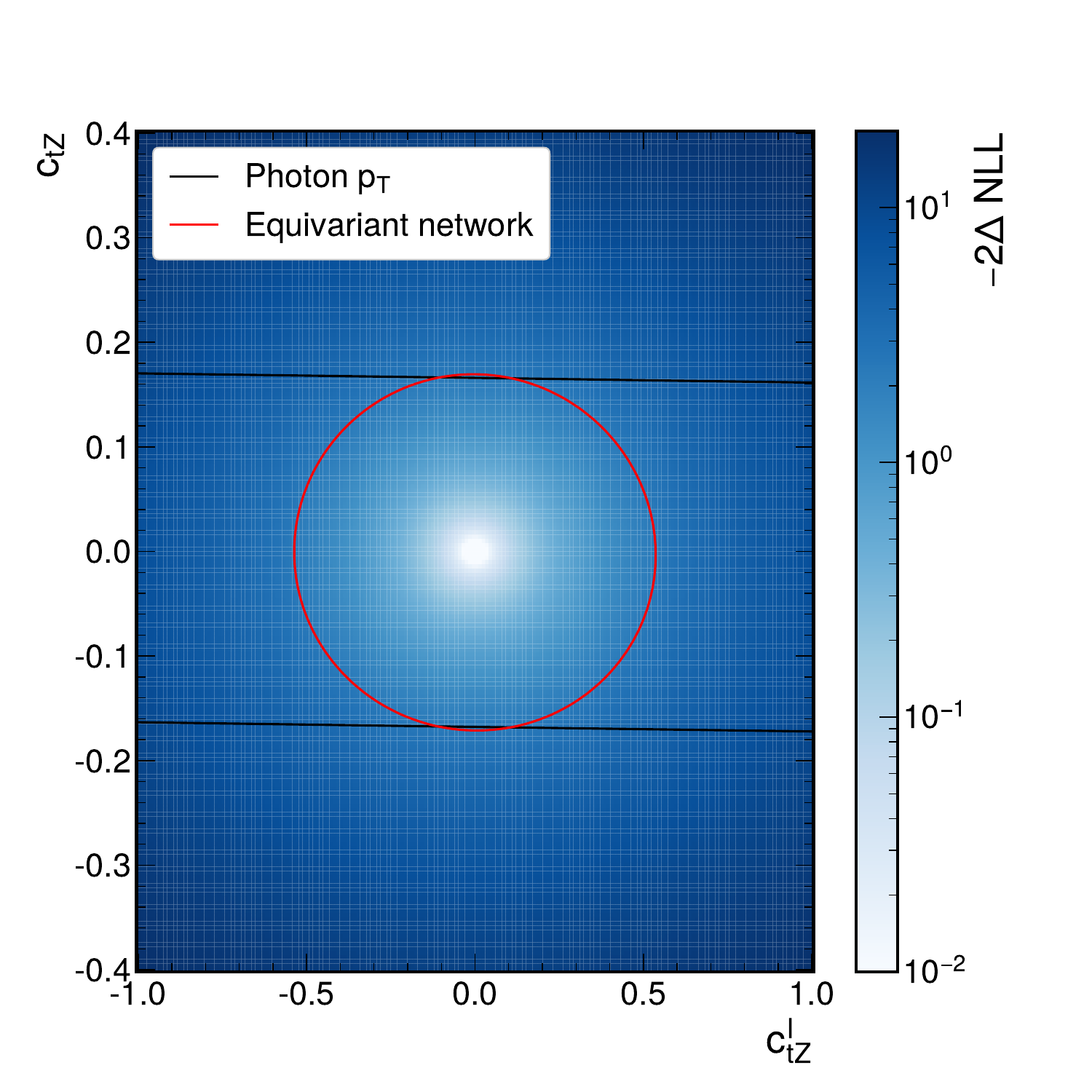}
\includegraphics[width=0.32\textwidth]{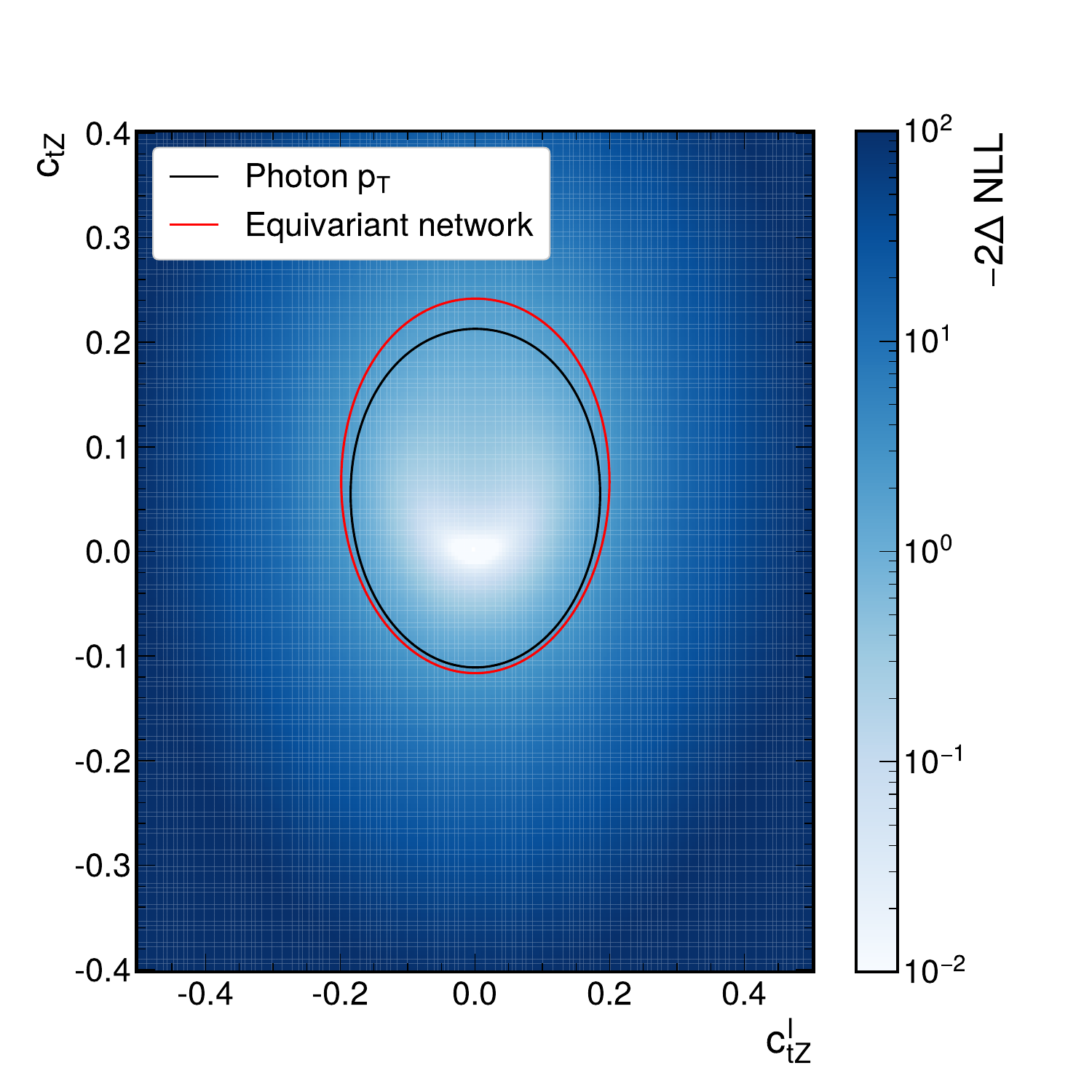}
\includegraphics[width=0.32\textwidth]{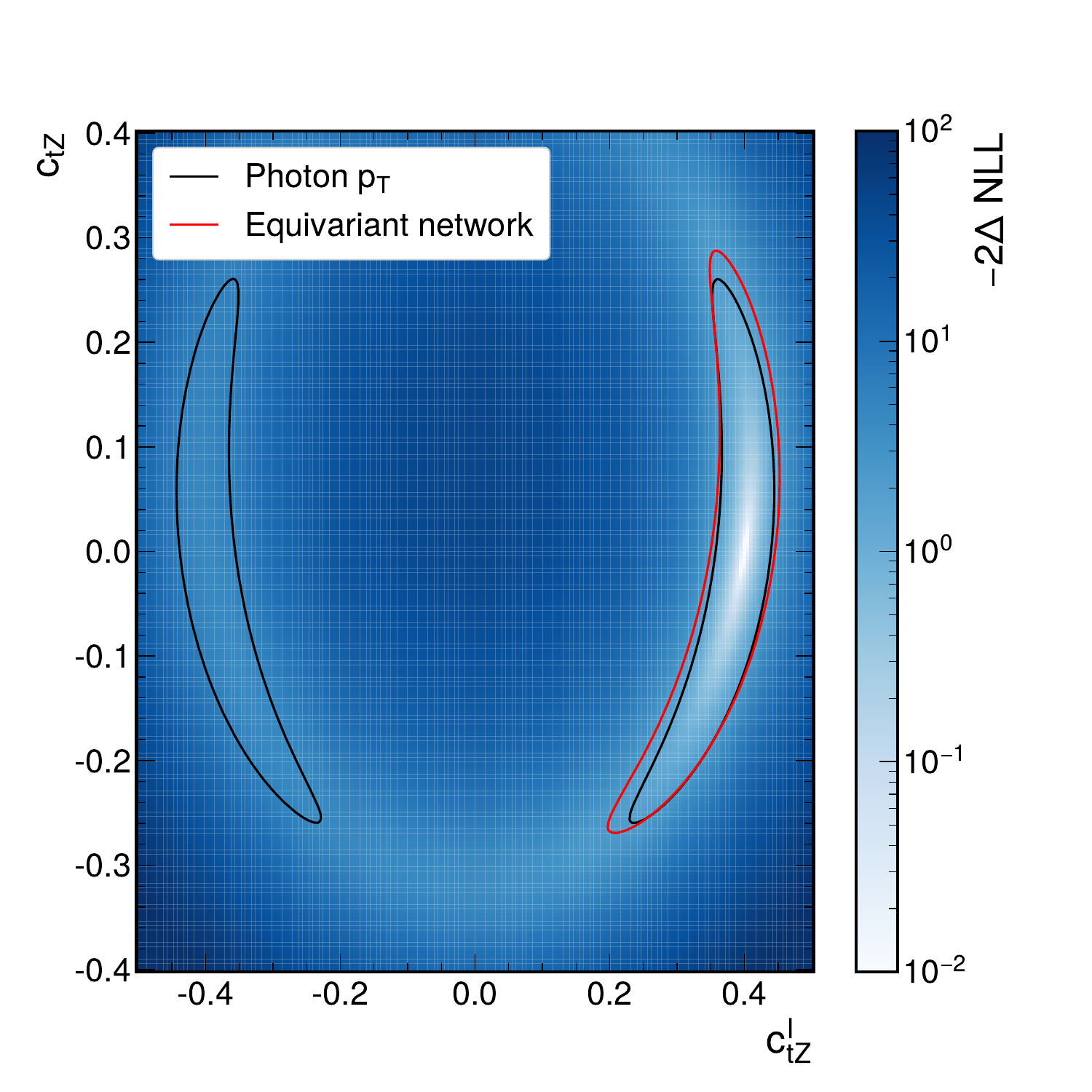}
\caption{Exclusion contours as a function of $c_{tZ}$ and $c_{tZ}^I$ for and analysis based on the photon \pt (red) and for an approach based on the distribution of the score of our algorithm (black), assuming a true value of $(c_{tZ}=0,c_{tZ}^I=0)$ and only linear contributions from the two operators (left), assuming a true value $(c_{tZ}=0,c_{tZ}^I=0)$ and including both linear and quadratic contributions, and assuming a true value $(c_{tZ}=0,c_{tZ}^I=0.4)$ also including linear and quadratic contributions.}
\label{fig:tta_results}
\end{figure*}

\subsection{CP violation in \texorpdfstring{$\ttA$}{ttG} production} 
\label{sec:tta} 

We consider \ttA production in final states where one of the top (anti)quarks decays leptonically and the other hadronically. In this case, we consider the effect of the $c_{tZ}$ and $c_{tZ}^I$ operators, which are \CP-even and odd, respectively. The effect of these operators in \ttA has been studied by the ATLAS and CMS Collaborations~\cite{CMS:2022lmh,ATLAS:2024hmk}, relying on the spectrum of the reconstructed photon \pt. In this section, we show that the sensitivity of such analyses can be significantly be improved by using \CP-odd observables constructed with our algorithm.

We consider events with a photon with $\pt > 50$ GeV, one lepton with $\pt > 25$ GeV and at least four jets with $\pt > 30$ GeV, out of which one \Pb-tagged. All of these objects are required to have $|\eta| < 2.5$. As input variables we use the three-momentum of the reconstructed photon $\overrightarrow{p_{\gamma}}$; the three-momentum of the reconstructed lepton $\overrightarrow{p_{\ell}}$ and the sign of its charge $Q_{\ell}$; the three-momenta of the four leading selected jets $\overrightarrow{p_{b_1}}$, $\overrightarrow{p_{b_2}}$, $\overrightarrow{p_{j_1}}$, and $\overrightarrow{p_{j_2}}$; and the the missing transverse momentum vector $\overrightarrow{ p_{\mathrm{T}}^{\mathrm{miss}}}$. As the jets $\Pb_1$ and $\Pb_2$, we pick the two leading \Pb-tagged jets and, if only one \Pb-tagged jet is present in the event, we pick that one together with the leading non-\Pb-tagged jet. We label the other two selected jets as $j_1$ and $j_2$. The indices 1 and 2 are chosen at random, as we did in the \ttbar measurement.  We choose $h_{\CP}$ as:

\begin{align}
    h_{\CP}\left( \overrightarrow{p_{\gamma}}, \overrightarrow{p_{\ell}},  Q_{\ell}, \overrightarrow{p_{b_1}}, \overrightarrow{p_{b_2}}, \overrightarrow{p_{j_1}}, \overrightarrow{p_{j_2}} \right) = \nonumber \\
    \left( -\overrightarrow{p_{\gamma}}, -\overrightarrow{p_{\ell}},  -Q_{\ell}, -\overrightarrow{p_{b_2}}, -\overrightarrow{p_{b_1}}, -\overrightarrow{p_{j_2}}, -\overrightarrow{p_{j_1}} \right)\,.
\end{align}

We then compare two analysis strategies, both of them classifying events in bins of a given distribution. For one of the strategies, we classify events as a function of the photon \pt, while, in the other, we classify events based on the output score of our equivariant algorithm. We build a likelihood model similar to the one in Eq.~\ref{eq:likelihood_cw}, where $\mu_j$ is now a function of $c_{tZ}^I$ and $c_{tZ}$, and we use it to draw exclusion contours in the $(c_{tZ},c_{tZ}^I)$ plane under different assumptions in Fig.~\ref{fig:tta_results}. Firstly, we show that, for an expected value of $(c_{tZ}=0,c_{tZ}^I=0)$ and taking into account only linear contributions of these two operators, our approach constrains $c_{tZ}$ equally well to the strategy based on the photon \pt. In addition, our approach is able to significantly constrain $c_{tZ}^I$ in this scenario, while the approach based on the photon \pt is not sensitive to this contribution, as it is a \CP-even observable. 

We also show results in which quadratic contributions are included. When the true value lies on $(c_{tZ}=0,c_{tZ}^I=0)$, the sensitivity of the two approaches is very similar, as the SMEFT contribution is dominated by the quadratic term. We remark that our approach, despite not being optimized to be sensitive to this contribution, performs nevertheless equally well. In addition, when the true value lies on $(c_{tZ}=0,c_{tZ}^I=0.4)$ the photon \pt approach does not provide sensitivity to the sign of $c_{tZ}^I$, showing a bimodal structure in the exclusion contours. In contrast, our approach is sensitive to the linear contributions, which translates into sensitivity to the sign of this operator. We conclude that our approach is superior to the state-of-the-art methodology in this channel.

\section{Conclusion} 
\label{sec:conclusions} 

We have developed an algorithm to construct optimal observables that transform equivariantly with the \CP symmetry, by introducing an inductive bias through the use of equivariant neural networks with respect to the $\mathds{Z}_2$ symmetry group. We have showcased the performance of our algorithm using realistic simulations of \ttbar, \WZ, and \ttA events, using it to produce \CP-odd observables that respect this symmetry property, regardless of the convergence of the training. We have also shown that the developed algorithm shows a faster numerical convergence of the method, requiring, in the presented benchmarks, between 40 and 300\% fewer epochs than a nonequivariant algorithm to be trained. We have also show that the observables produced with the method improve the state-of-the-art methodology used for \CP-violating searches in \WZ and \ttA production.

\section*{Ackowledgements} 
\label{sec:acknowledgements} 

P.V.’s work was supported by the ``Ramón y Cajal'' program under Project No. RYC2021-033305-I funded by MCIN/AEI/10.13039/501100011033 and by the European Union NextGenerationEU/PRTR.

P.V. and C.R.A. acknowledge the computer resources at Artemisa, funded by the European Union ERDF and Comunitat Valenciana as well as the technical support provided by the Instituto de Fisica Corpuscular, IFIC (CSIC-UV).

\bibliographystyle{lucas_unsrt}
\bibliography{biblio}% Produces the bibliography via BibTeX.

\end{document}